\documentclass[10pt]{iopart}

\newcommand{\bra}[1]{\langle #1 |}
\newcommand{\ket}[1]{| #1 \rangle}

\newcommand{\ua}[0]{ \uparrow}
\newcommand{\da}[0]{ \downarrow}

\newcommand{\md}[0]{ {\rm media}}
\newcommand{\op}[0]{ {\rm op}}
\newcommand{\tot}[0]{ {\rm tot}}
\newcommand{\bbra}[1]{\bigl \langle #1 \bigl|}
\newcommand{\bket}[1]{\bigl | #1 \bigr \rangle}
\newcommand{\eqref}[1]{(\ref{#1})}

\usepackage{graphicx}
\usepackage{subfigure}
\usepackage{color}
\begin{document}

\title[  ]{Role of external fields in enhancing long-distance entanglement at finite temperatures}

\author{Tomotaka Kuwahara}

\address{Department of Physics, The University of Tokyo, Komaba, Meguro, Tokyo 153-8505}
\ead{tomotaka@iis.u-tokyo.ac.jp}

\begin{abstract}
We  investigate the end-to-end entanglement  of a general $XYZ$-spin chain at the non-zero temperatures.
The entanglement usually vanishes at a certain critical temperature~$T_c$, but external fields can make $T_c$ higher.
We obtain a general statement on the increase of the critical temperature~$T_c$ by the external fields.
We prove that if the two end spins are separated by two spins or more, the critical temperature cannot be higher than a certain finite temperature $\bar{T}_c$ ($T_c\le\bar{T}_c$), that is, the entanglement must vanish above the temperature~$\bar{T}_c$ for any values of the external fields.
On the other hand, if the two end spins are separated by one spin, the entanglement maximized by the external fields exhibits a  power law decay of the temperature, being finite at any temperatures.
In order to demonstrate the former case, we numerically calculate the temperature $\bar{T}_c$ in  \textit{XX} and \textit{XY} four-spin chains.
We find that the temperature~$\bar{T}_c$ shows qualitatively different behavior, depending on the conservation of the angular momentum in the \textit{z} direction.
\end{abstract}

\maketitle

\section{Introduction}
Quantum entanglement has played an important role in various fields~\cite{Nielsen, Amico, horodecki,Area_law} and hence many researchers have investigated the fundamental properties of the entanglement.
In particular, the entanglement generation and distribution have been two of the most important targets~\cite{XWang,Asoudeh,Nakata,modular,Entropy1,Entropy2,Entropy3,Entropy4,Entropy5,Entropy6}.
The entanglement between two systems is typically generated by an interaction between them, but the amount of the entanglement is not simply determined because it strongly depends on the interaction and the external fields on these systems.

\begin{figure}
\centering

\includegraphics[clip, scale=0.6]{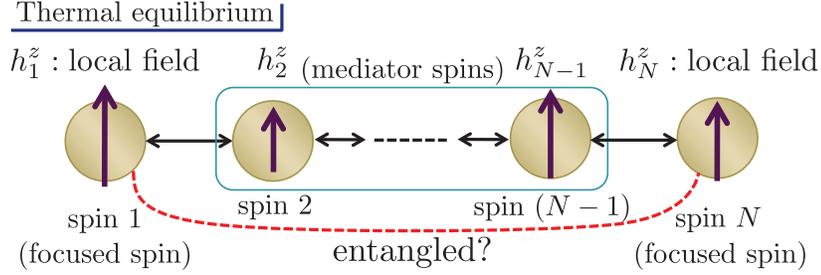}

\caption{Schematic picture of the system which we consider. We define the `focused' spins as the spins between which we mainly consider the entanglement, the `mediator' system as the quantum system which mediates  the indirect interaction and the `local fields' as the external fields on the focused spins.  Here, we define the spins~1 and \textit{N} as the focused spins and the other spins $2\le i \le N-1$ as the mediator spins.
We assume that we can tune the local fields on the spins 1 and \textit{N}, while the other fields $\{h_i^z\}_{i=2}^{N-1}$ are arbitrary but fixed.}
\label{fig:spin_chians_theorem2}
\end{figure}

We here aim to obtain general properties on the generation of the entanglement between two spins which indirectly interact with each other through another quantum system, which we refer to as a `mediator' system.
Hereafter, we refer to the two spins between which we consider the entanglement as  `focused spins' and refer to the external fields on the focused spins as `local fields' (Fig.~\ref{fig:spin_chians_theorem2}).
In the following, we mean by the field-spin interaction an operator on one spin such as $h^x \sigma^x+h^y \sigma^y+h^z \sigma^z$, where $\{\sigma^\xi\}_{\xi=x,y,z}$ are the Pauli matrices. 

When two spins indirectly interact with each other, we mainly have to consider the following three factors; (i) the thermal fluctuation, (ii) the entanglement with the mediator system, and (iii) the local fields on the focused spins.
These factors complicate the mechanism of the entanglement generation.
The first two factors generally decrease the purity of the focused spins and thereby contribute to the entanglement destruction, whereas the third factor can enhance the entanglement generation, depending on the situation.

Previous studies have partly clarified the effects of the above three factors on the entanglement generation between two spins~\cite{Fujinaga,Fardin,Plastina2,Plastina3,Plastina4,Bose2,JHide}.
For example, in the low-temperature limit, it has been shown that the entanglement can be generated over long distance in a spin chain, that is, a large mediator system~\cite{LDE1,LDE2,Plastina,LDE3,LDE4,LDE5,LDE6,LDE7,LDE8,LDE9,LDE10}; such an entanglement is referred to as `the long-distance entanglement' and the general conditions for the long-distance entanglement has been obtained in Ref.~\cite{Kuwahara3}.

At finite temperatures, on the other hand, it is well known~\cite{Arnesen} that the entanglement vanishes above a certain critical temperature $T_c$.
Several studies~\cite{Bose1,Kamta,Kuwahara} have also pointed out that adding the local fields on the focused two spins can enhance the entanglement generation and increase the critical temperature.
Indeed, it has been discovered~\cite{Kuwahara} that in two-spin systems we can make the critical temperature infinite by applying appropriate local fields; in such cases, the entanglement decays as $1/(T\ln T)$ in the optimizing local fields.
In the high-temperature limit, therefore, the entanglement remains but is infinitesimally small with the local fields.

When the number of spins is more than two, we may also achieve the increase of the critical temperature by adding the local fields.
In fact, there are no analytical studies on the general properties of the increase of the critical temperature in systems with more than two spins.
In such systems, there are two possibilities in the high-temperature limit under the condition that we tune the local fields arbitrarily: the entanglement completely vanishes or it infinitesimally remains. 
It is important to know whether the entanglement is exactly zero or not because we can concentrate the entanglement if it has a non-zero value~\cite{concentrate_Benett}.

In our research, we consider the end-to-end entanglement  of a general $XYZ$-spin chain and investigate how high the critical temperature becomes in the presence of the local fields (Fig.~\ref{fig:spin_chians_theorem2}~(b)).
In this case, the critical temperature is a function of the local fields on the two focused spins, namely $T_c(h_1,h_N)$.
We want to know the upper limit of the critical temperature, namely ${\displaystyle \max_{h_1,h_N} \bigl[T_c(h_1,h_N)\bigr]}$.
In order to study it, we consider the maximum value of the end-to-end under the condition that we arbitrarily tune the local fields only on the two end spins.
We regard the other parameters such as the interaction parameters and the external fields on the mediator system as fixed. 
In some cases, such a `maximized entanglement'  is equal to zero above a certain temperature, which means that the entanglement generation is impossible for any values of the local fields.
In other words, the critical temperature cannot be increased beyond such a temperature.
This temperature gives the upper limit of the critical temperature ${\displaystyle \max_{h_1,h_N} \bigl[T_c(h_1,h_N)\bigr]}$, which we define as $\bar{T}_c$ in distinction from the normal critical temperature $T_c$.
For example, in two-spin systems, $\bar{T}_c$ is infinite because the entanglement never vanishes in appropriate local fields.

Our main result is the following: at high temperatures, we prove that the maximized entanglement is always equal to zero between the two spins which are separated by two or more spins $(N\ge 4)$. In other words, above a finite temperature $\bar{T}_c$, we can never generate the entanglement between spins far apart for any values of the local fields. 
Our result shows that there are limitations to the increase of the critical temperature. 
The reason is the following. 
The local fields enhance the entanglement generation mainly because they increase the purity of the focused spins.
It means that the local fields suppress the effects of thermal fluctuation and the entanglement with the mediator system. 
At the same time, too strong local fields destroy the correlation between the two spins, bringing them close to the product state.
For two and three spin chains $(N\le 3)$, the suppression of the purity is more dominant than the decoupling effect, and hence the entanglement can survive at high temperatures in appropriate local fields. 
For more than three spins $(N\ge 4)$, the decoupling by the local fields is more dominant than the suppression of the purity and hence we cannot keep non-zero entanglement for any values of the local fields above the temperature~$\bar{T}_c$.
The boundary between the two cases of $N\le3$ and $N\ge4$ arises because of the following reason.  
In the limit $|h_1|\to \infty$ and $|h_N|\to \infty$, the positive contribution to the entanglement is roughly given by $\langle \sigma_2\sigma_{N-1}\rangle/(|h_1||h_N|)$, whereas the negative contribution is given by $1/(|h_1||h_N|)$, where $\langle \sigma_2\sigma_{N-1}\rangle$ is the correlation between the spins~2 and $N-1$.
The term $\langle \sigma_2\sigma_{N-1}\rangle$ decays as $\beta^{N-3}$ in the high temperature limit $\beta\to0$. 
In the case $N\ge 4$, then, the positive contribution to the entanglement decays more rapidly than the negative one.

This paper consists of the following sections:  in Section~2, we state the problem specifically and give some definitions; in Section~3, we give a general theorem on the entanglement generation which is applied to any spin chains; in Section~4, we show the numerical and analytical results on the maximization of the entanglement in four-spin chains; in Section~5, discussion concludes the paper.

%

%

%%%%%%%%%%%%%%%%%%%%%%%%%%%%%%%%%%%%%%%%%%%%%%%%%%%%%%%%%%%%%%%%%%%%%%%%%%%%%%%%%%%%%%%%%%%%%%%%%%%%%%%%%%%%%%%%
%%%%%%%%%%%%%%%%%%%%%%%%%%%%%%%%%%%%%%%%%%%%%%%%%%%%%%%%%%%%%%%%%%%%%%%%%%%%%%%%%%%%%%%%%%%%%%%%%%%%%%%%%%%%%%%%
%%%%%%%%%%%%%%%%%%%%%%%%%%%%%%%%%%%%%%%%%%%%%%%%%%%%%%%%%%%%%%%%%%%%%%%%%%%%%%%%%%%%%%%%%%%%%%%%%%%%%%%%%%%%%%%%
%%%%%%%%%%%%%%%%%%%%%%%%%%%%%%%%%%%%%%%%%%%%%%%%%%%%%%%%%%%%%%%%%%%%%%%%%%%%%%%%%%%%%%%%%%%%%%%%%%%%%%%%%%%%%%%%
%%%%%%%%%%%%%%%%%%%%%%%%%%%%%%%%%%%%%%%%%%%%%%%%%%%%%%%%%%%%%%%%%%%%%%%%%%%%%%%%%%%%%%%%%%%%%%%%%%%%%%%%%%%%%%%%
%%%%%%%%%%%%%%%%%%%%%%%%%%%%%%%%%%%%%%%%%%%%%%%%%%%%%%%%%%%%%%%%%%%%%%%%%%%%%%%%%%%%%%%%%%%%%%%%%%%%%%%%%%%%%%%%
%%%%%%%%%%%%%%%%%%%%%%%%%%%%%%%%%%%%%%%%%%%%%%%%%%%%%%%%%%%%%%%%%%%%%%%%%%%%%%%%%%%%%%%%%%%%%%%%%%%%%%%%%%%%%%%%
%%%%%%%%%%%%%%%%%%%%%%%%%%%%%%%%%%%%%%%%%%%%%%%%%%%%%%%%%%%%%%%%%%%%%%%%%%%%%%%%%%%%%%%%%%%%%%%%%%%%%%%%%%%%%%%%

\section{Statement of the problem}

First, we formulate the framework of the entanglement maximization problem and describe conditions.
We consider a general $XYZ$ $N$-spin chain with external fields in the \textit{z} direction. 
The most general form of the Hamiltonian of this system is given as follows:
\begin{eqnarray}
H_{{\rm  tot}}=&\sum_{i=1}^{N-1} ( J_i^x \sigma_i^x   \sigma_{i+1}^x + J_i^y \sigma_i^y   \sigma_{i+1}^y+ J_i^z \sigma_i^z   \sigma_{i+1}^z)  + \sum_{i=1}^N h_i^z \sigma_i^z ,\label{Fundamental_Hamiltonian_Setup}
\end{eqnarray} 
where ${\{\sigma_i^i\}_{i=x,y,z}}$ are the Pauli matrices and we adopt the free boundary conditions.

We hereafter consider the entanglement between the spins $1$ and $N$ at the ends of the chain (Fig.~\ref{fig:spin_chians_theorem2}~(b)).
We refer to these two spins as the `focused spins' and refer to the external fields $h_1^z$ and $h_N^z$ on the two spins as the `local fields'.
We refer to the other spins $2\le i \le N-1$ as the `mediator spins'. 
We want to obtain the maximum value of the thermal entanglement between the focused spins $1$ and $N$ by tuning the local fields $h_1^z$ and $h_N^z$ at a fixed temperature.
We also fix all the other parameters, namely, $\{J_i^x,J_i^y,J_i^z\}$ for $1\le i \le N-1$ and $\{h_i^z\}$ for $2\le i \le N-1$.
We refer to the maximizing values of the local fields $h_1^z$ and $h_N^z$ as $h_{1\op}$ and $h_{N\op}$.

Note that the maximizing local fields $h_{1\op}$ and $h_{N\op}$ generally depend on the temperature $T$, or on the inverse temperature $\beta=1/(kT)$ with $k$ the Boltzmann constant. 
This is because we tune the local fields at a fixed temperature $\beta$.
Let us then define the high-temperature limit, in which we mostly develop the argument.
In our high-temperature limit, we make $\beta||H_{{\rm  tot}}-h_1^z \sigma_1^z-h_N^z \sigma_N^z||$ tend to zero, where $||\cdots||$ denotes the matrix norm.
On the other hand, we let the maximizing local fields depend on $\beta$ as we take the limit $\beta\rightarrow 0$, and we sometimes denote them as $h_{1\op}(\beta)$ and $h_{N\op}(\beta)$.
Hence, $\beta h_{1\op}(\beta)$ and $\beta h_{N\op}(\beta)$ may even diverge in our high-temperature limit.

The density matrix of the total system in thermal equilibrium is 
\begin{eqnarray}
\rho_{{\rm  tot}} &= \frac{e^{-\beta H_{{\rm  tot}}}}{Z_{{\rm  tot}}},     \label{density matrix}
\end{eqnarray}
where ${Z_{{\rm  tot}} ={\rm  tr}( e^{-\beta H_{{\rm  tot}}}) }$ is the partition function.
The density matrix of the focused spins $1$ and $N$ is
\begin{eqnarray}
\rho_{1N} &= \tr_{1N}\rho_{{\rm  tot}},
\end{eqnarray}
where $\tr_{1N}$ denotes the trace operation on the spins \textit{except} the focused  spins~$1$ and $N$.
For the present system \eqref{Fundamental_Hamiltonian_Setup}, the general form of the density matrix $\rho_{1N}$ is given by:
\begin{eqnarray}
\rho_{1N}= 
\left( 
\begin{array}{cccc}
p_{\ua\ua}&0&0&F_2 \\
0&p_{\ua\da}&F_1&0\\
0&F_1&p_{\da\ua}&0\\
F_2&0&0&p_{\da\da}                
\end{array}    \label{denmat_paper3}
\right) 
\end{eqnarray} 
in the basis of the eigenstates of $\sigma_1^z\otimes\sigma_N^z$, where $p_{\ua\ua}$, $p_{\ua\da}$, $p_{\da\ua}$, $p_{\da\da}$, $F_1$ and $F_2$ are real numbers.
We have $F_2=0$ when $J_i^x=J_i^y$, for $1\le i \le N-1$, in particular.

In order to quantify the entanglement, we here adopt the concurrence~\cite{Wootter}, which is most commonly used as an entanglement measure. The concurrence $C(\rho_{1N})$ is defined as follows:
\begin{eqnarray} 
C(\rho_{1N}) \equiv \max(\lambda_1-\lambda_2-\lambda_3-\lambda_4,0)  ,
\end{eqnarray} 
where $\{\lambda_n\}_{n=1}^4$ are the eigenvalues of the matrix
$
\sqrt{\rho_{1N} (\sigma_1^y\otimes \sigma_N^y) \rho_{1N}^{\ast}(\sigma_1^y\otimes \sigma_N^y) }
$
in the non-ascending order $\lambda_1\ge\lambda_2\ge\lambda_3\ge\lambda_4$.
For density matrices of the form \eqref{denmat_paper3}, the concurrence $C(\rho_{1N})$ is reduced to the simpler form
\begin{eqnarray} 
C(\rho_{1N}) =2 \max(|F_1|-\sqrt{p_{\ua\ua}p_{\da\da}},|F_2|-\sqrt{p_{\ua\da}p_{\da\ua}},0) .   \label{conexpression}
\end{eqnarray} 
Then, the necessary and sufficient condition for the existence of the entanglement is given by
\begin{eqnarray} 
\max(|F_1|-\sqrt{p_{\ua\ua}p_{\da\da}},|F_2|-\sqrt{p_{\ua\da}p_{\da\ua}} ) >0 .   \label{sufneccond1_paper3}
\end{eqnarray} 
Thus, the present entanglement optimization problem for the spin pair $(1,N)$ is equivalent to finding the values of ${\{h_1^z,h_N^z\}}$ which maximize $C(\rho_{1N})$ with the parameters $\{J_i^x,J_i^y,J_i^z\}$ ($1\le i \le N-1$), $\{h_i^z\}$ ($2\le i \le N-1$), and $\beta$ all fixed.
We denote the maximized entanglement by $C_{\rm op}$:
\begin{eqnarray} 
C_{\rm op} &= \max_{h_1^z,h_N^z} \bigl[C(\rho_{1N}) \bigr]=C(\rho_{1N}) \bigl |_{h_1^z=h_{1{\rm op}},h_N^z=h_{N{\rm op}}} .
\end{eqnarray}

%Before presenting our main results on the entanglement optimization,
%we prove the following Lemma~1 to simplify the present entanglement optimization problem. 
%%%%%%%%%%%%%%%%%%%%%%%%%%%%%%%%%%%%%%%%%%%%%%%%%%%%%%%%%%%%%%%%%%%%%%%%%%%%%%%%%%%%%%%%%%%%%%%%%%%%%%%%%%%%%%%%
%%%%%%%%%%%%%%%%%%%%%%%%%%%%%%%%%%%%%%%%%%%%%%%%%%%%%%%%%%%%%%%%%%%%%%%%%%%%%%%%%%%%%%%%%%%%%%%%%%%%%%%%%%%%%%%%
%%%%%%%%%%%%%%%%%%%%%%%%%%%%%%%%%%%%%%%%%%%%%%%%%%%%%%%%%%%%%%%%%%%%%%%%%%%%%%%%%%%%%%%%%%%%%%%%%%%%%%%%%%%%%%%%
%%%%%%%%%%%%%%%%%%%%%%%%%%%%%%%%%%%%%%%%%%%%%%%%%%%%%%%%%%%%%%%%%%%%%%%%%%%%%%%%%%%%%%%%%%%%%%%%%%%%%%%%%%%%%%%%
%%%%%%%%%%%%%%%%%%%%%%%%%%%%%%%%%%%%%%%%%%%%%%%%%%%%%%%%%%%%%%%%%%%%%%%%%%%%%%%%%%%%%%%%%%%%%%%%%%%%%%%%%%%%%%%%
%%%%%%%%%%%%%%%%%%%%%%%%%%%%%%%%%%%%%%%%%%%%%%%%%%%%%%%%%%%%%%%%%%%%%%%%%%%%%%%%%%%%%%%%%%%%%%%%%%%%%%%%%%%%%%%%
%%%%%%%%%%%%%%%%%%%%%%%%%%%%%%%%%%%%%%%%%%%%%%%%%%%%%%%%%%%%%%%%%%%%%%%%%%%%%%%%%%%%%%%%%%%%%%%%%%%%%%%%%%%%%%%%
%%%%%%%%%%%%%%%%%%%%%%%%%%%%%%%%%%%%%%%%%%%%%%%%%%%%%%%%%%%%%%%%%%%%%%%%%%%%%%%%%%%%%%%%%%%%%%%%%%%%%%%%%%%%%%%%

 \section{General theorems on entanglement generation}
In the previous section, we formulated the entanglement optimization problem.
In the present section, we introduce the main theorem on entanglement generation by optimizing the local fields.

\textit{Theorem~1}.  
Let us consider the \textit{XYZ} chain \eqref{Fundamental_Hamiltonian_Setup} with $\{J_1^x,J_1^y,J_1^z\}=\{J_{N-1}^x,J_{N-1}^y,J_{N-1}^z\}=\{J^x,J^y,J^z\}$.
We tune the local fields $h_1^z$ and $h_N^z$, while the external fields on the mediator spins $\{h_i^z\}_{i=2}^{N-1}$ are arbitrary but fixed, thus obtaining the maximized entanglement $C_{\rm op}$.
There is a certain temperature $\bar{T}_c$ above which the maximized entanglement $C_{\rm op}$ between the focused spins 1 and \textit{N} is exactly zero if they are separated by two or more spins ($N\ge 4$).
In other words, we cannot generate the entanglement for any values of the local fields above this temperature $\bar{T}_c$ for $N\ge4$.

\textit{Comments}. This temperature gives an upper bound of the critical temperature $T_c$ with respect to the parameters $h_1^z$ and $h_N^z$, namely ${\displaystyle \bar{T}_c=\max_{h_1^z,h_N^z} \bigl[T_c(h_1^z,h_N^z)\bigr]}$.
In other words, for  $T< \bar{T}_c$, we can always generate the entanglement by choosing the local fields appropriately.
We say $\bar{T}_c=\infty$ if we can generate the entanglement at any temperatures.

In this theorem, we discuss the case of $\{J_1^x,J_1^y,J_1^z\}=\{J_{N-1}^x,J_{N-1}^y,J_{N-1}^z\}=\{J^x,J^y,J^z\}$, but the case of $\{J_1^x,J_1^y,J_1^z\}\neq\{J_{N-1}^x,J_{N-1}^y,J_{N-1}^z\}$ can be also proved by extending the following proof; we only have to repeat the same calculation by changing $H_{{\rm  couple}}$ in Eq.~\eqref{Definition_H_media_couple} below accordingly.

%%%%%%%%%%%%%%%%%%%%%%%%%%%%%%%%%%%%%%%%%%%%%%%%%%%%%%%%%%%%%%%%%%%%%%%%%%%%%%%%%%%%%%%%%%%%%%%%%%%%%%%%%%%%%%%%
%%%%%%%%%%%%%%%%%%%%%%%%%%%%%%%%%%%%%%%%%%%%%%%%%%%%%%%%%%%%%%%%%%%%%%%%%%%%%%%%%%%%%%%%%%%%%%%%%%%%%%%%%%%%%%%%
%%%%%%%%%%%%%%%%%%%%%%%%%%%%%%%%%%%%%%%%%%%%%%%%%%%%%%%%%%%%%%%%%%%%%%%%%%%%%%%%%%%%%%%%%%%%%%%%%%%%%%%%%%%%%%%%
%%%%%%%%%%%%%%%%%%%%%%%%%%%%%%%%%%%%%%%%%%%%%%%%%%%%%%%%%%%%%%%%%%%%%%%%%%%%%%%%%%%%%%%%%%%%%%%%%%%%%%%%%%%%%%%%

\textit{Proof}. 
In order to prove this theorem, it is enough to show that in \textit{the high temperature limit $\beta \rightarrow 0$} we have
\begin{eqnarray}
{\rm  max}(F_1^2- p_{\ua\ua} p_{\da\da} , F_2^2 - p_{\ua\da} p_{\da\ua} )  < 0     \label{Prove_this_inequality}
\end{eqnarray} 
after the maximization of the left-hand side with respect to $h_1^z$ and $h_N^z$, where $\{F_1,F_2,p_{\ua\ua}, p_{\ua\da} , p_{\da\ua} ,p_{\da\da}\}$ are the elements of the density matrix defined in Eq.~\eqref{denmat_paper3}.
Then, Eq.~\eqref{conexpression} yields the exactly zero concurrence $C_{\rm op}=0$ in the limit $\beta\rightarrow 0$.
Since the system \eqref{Fundamental_Hamiltonian_Setup} has a finite number of degrees of freedom, the elements $\{F_1,F_2,p_{\ua\ua}, p_{\ua\da} , p_{\da\ua} ,p_{\da\da}\}$ must be analytic as a function of $\beta$.
Therefore, there can be a finite value of $\beta$ at which ${\rm  max}(F_1^2- p_{\ua\ua} p_{\da\da} , F_2^2 - p_{\ua\da} p_{\da\ua} ) =0$ after the maximization.
This gives the temperature~$\bar{T}_c$. 
Note that the elements of $\rho_{1N}$ here are functions of $h_{1\op}(\beta)$, $h_{N\op}(\beta)$ and $\beta$.
The increases of the elements $\{p_{\ua\ua}p_{\da\da},p_{\ua\da}p_{\da\ua}\}$ contribute to the entanglement destruction as can be seen in Eq.~\eqref{Prove_this_inequality} but they can be decreased by the local fields mainly because of the purification by the local fields.
On the other hand, the decoupling effect of the local fields also decreases the elements $\{F_1^2, F_2^2\}$ and thereby contributes to the entanglement destruction according to Eq.~\eqref{Prove_this_inequality}.
We here show that in the case $N\ge4$, the elements $\{F_1^2, F_2^2\}$ always decay more rapidly than the elements $\{p_{\ua\ua}p_{\da\da},p_{\ua\da}p_{\da\ua}\}$ as the temperature increases.

For the proof of Eq.~(9), we focus on the temperature dependence of the local fields $h_{1{\rm op}}(\beta)$ and $h_{N{\rm op}}(\beta)$.
Note that the local fields which maximize the entanglement $C(\rho_{1N})$ depend on the temperature.
Let us define that the $\beta$ dependence of $h_{1\op}(\beta)$ and $h_{N\op}(\beta)$ are, respectively, of order of
\begin{eqnarray}
\beta^{-\kappa_1} \  {\rm and} \  \beta^{-\kappa_N}      \label{definition_kappa1N}
\end{eqnarray}  
in the limit $\beta\rightarrow0$, where $\kappa_1$ and $\kappa_N$ are real numbers.
By the phrase ``$f(\beta)$ is of order of $\beta^\kappa$,'' we mean 
\begin{eqnarray}
\lim_{\beta\to0}\log_\beta f(\beta)=\kappa.
\end{eqnarray}
If two functions $f(\beta)$ and $g(\beta)$ are of order of $\beta^\kappa$ and $\beta^{\kappa'}$ ($\kappa>\kappa'$) respectively, $f(\beta)$ decays more rapidly than $g(\beta)$ in the limit $\beta\rightarrow0$ and we have 
\begin{eqnarray}
f(\beta)-g(\beta)<0 .
\end{eqnarray}
In the following, we investigate and compare the leading orders of the elements $\{F_1^2, F_2^2\}$ and $\{p_{\uparrow\uparrow}p_{\downarrow\downarrow},p_{\uparrow\downarrow}p_{\downarrow\uparrow}\}$ and prove that for any values of $\kappa_1$ and $\kappa_N$ the inequality~(9) is satisfied.

We estimate the order of each element of $\{F_1^2, F_2^2\}$ and $\{p_{\ua\ua}p_{\da\da},p_{\ua\da}p_{\da\ua}\}$ in the following three cases:
\begin{itemize}
\item{Case~(a):} $\kappa_1<1$ and $\kappa_N<1$.
\item{Case~(b):} $\kappa_1\ge \kappa_N$, $\kappa_1\ge1$ and $\kappa_N>0$; or $\kappa_N\ge \kappa_1$, $\kappa_1>0$ and $\kappa_N\ge1$.
\item{Case~(c):} $\kappa_1\ge1$ and $\kappa_N\le0$; or $\kappa_1\le0$ and $\kappa_N\ge1$.
\end{itemize}
Notice that the three cases cover the entire space of  $\kappa_1$ and $\kappa_N$.

%%%%%%%%%%%%%%%%%%%%%%%%%%%%%%%%%%%%%%%%%%%%%%%%%%%%%%%%%%%%%%%%%%%%%%%%%%%%%%%%%%%%%%%%%%%%%%%%%%%%%%%%%%%%%%%%
%%%%%%%%%%%%%%%%%%%%%%%%%%%%%%%%%%%%%%%%%%%%%%%%%%%%%%%%%%%%%%%%%%%%%%%%%%%%%%%%%%%%%%%%%%%%%%%%%%%%%%%%%%%%%%%%

\textit{Case~(a)}. 
In this case, we prove the inequality~\eqref{Prove_this_inequality} by utilizing a necessary condition for the existence of the entanglement~\cite{Zyczkowski}, which is
\begin{eqnarray}
\mathop{\mathrm{tr}}\rho_{1N}^2\ge \frac{1}{3}.
\end{eqnarray}
In the case (a), $\beta h_{1\op}$ and $\beta h_{N\op}$ are of order of $\beta^{1-\kappa_1}$ and $\beta^{1-\kappa_N}$, respectively, and approach to zero in the high temperature limit $\beta \rightarrow 0$.
We then have $\beta||H_{{\rm  tot}}||\rightarrow0$ and therefore, the density matrix $\rho_{1N}$ becomes proportional to the identity matrix. We hence have
\begin{eqnarray}
\lim_{\beta\to0}\mathop{\mathrm{tr}}\rho_{1N}^2=\frac{1}{4}.
\end{eqnarray}
Therefore, in the case~(a), the entanglement between the spins $1$ and $N$ is exactly zero in the high-temperature limit.

%%%%%%%%%%%%%%%%%%%%%%%%%%%%%%%%%%%%%%%%%%%%%%%%%%%%%%%%%%%%%%%%%%%%%%%%%%%%%%%%%%%%%%%%%%%%%%%%%%%%%%%%%%%%%%%%
%%%%%%%%%%%%%%%%%%%%%%%%%%%%%%%%%%%%%%%%%%%%%%%%%%%%%%%%%%%%%%%%%%%%%%%%%%%%%%%%%%%%%%%%%%%%%%%%%%%%%%%%%%%%%%%%

\textit{Case~(b)}. 
To simplify the problem, we consider the case of $ h_{1\op},  h_{N\op}>0$, $\kappa_1\ge \kappa_N$, $\kappa_1\ge1$ and $\kappa_N>0$, but we can prove the other cases in the same way.
In addition, we consider the case where $h_{1\op}-  h_{N\op}$ is of order of     
\begin{eqnarray}
\beta^{-\tilde{\kappa}} \ {\rm with}\  \tilde{\kappa}>0   \label{Paraorders1_for_Appendix}
\end{eqnarray}
in the limit $\beta\rightarrow0$ in the following.
We discuss the case $\tilde{\kappa}\le0$ in~\ref{appendixE_paper3}. 
In order to obtain the inequality~\eqref{Prove_this_inequality}, we prove the following; $F_{1}$ and $F_{2}$ are of order of
\begin{eqnarray}
\beta^{\kappa_1+\kappa_N + \kappa}  \  {\rm  and} \  \beta^{\kappa_1+\kappa_N + \kappa'}\  {\rm or} \ {\rm  higher}  ,    \label{Paraorders1}
\end{eqnarray}
respectively, where
\begin{eqnarray}
\kappa=\min(\kappa_N,1)>0 \ {\rm  and}\  \kappa'=\min(\kappa_N,\tilde{\kappa},1)>0 .  \label{Definition_of_the_two_kappa}
\end{eqnarray}
On the other hand, $p_{\ua\ua}p_{\da\da}$ and $p_{\ua\da}p_{\da\ua}$ are both of order of
\begin{eqnarray}
\beta^{2\kappa_1+2\kappa_N} \  {\rm or} \ {\rm  lower} .  \label{Paraorders2}
\end{eqnarray} 
Then, the inequality~\eqref{Prove_this_inequality} is satisfied below a certain value of $\beta$.

In order to prove \eqref{Paraorders1} and \eqref{Paraorders2}, we separate the total Hamiltonian as follows:
\begin{eqnarray}
H_{{\rm  tot}} = H_1 +H_{{\rm  couple}}+  H_{{\rm  media}}+ H_N  ,
\end{eqnarray} 
where 
\begin{eqnarray}
\fl &H_1=h_{1\op} \sigma_1^z, \quad H_N =  h_{N\op} \sigma_N^z, \nonumber \\
\fl &H_{{\rm  couple}} = J^x \sigma_1^x   \sigma_{2}^x + J^y \sigma_1^y   \sigma_{2}^y + J^z   \sigma_1^z   \sigma_{2}^z  + J^x \sigma_{N-1}^x   \sigma_{N}^x + J^y \sigma_{N-1}^y   \sigma_{N}^y + J^z   \sigma_{N-1}^z   \sigma_{N}^z,   \nonumber \\
 \fl &\quad \quad \quad  =\frac{1}{2}  J\bigl[(1+\gamma) \sigma_1^x   \sigma_{2}^x + (1- \gamma) \sigma_1^y   \sigma_{2}^y\bigr] + J^z   \sigma_1^z   \sigma_{2}^z  \nonumber \\
\fl&\quad  \quad \quad +\frac{1}{2}  J\bigl[(1+\gamma) \sigma_{N-1}^x   \sigma_{N}^x + (1- \gamma) \sigma_{N-1}^y   \sigma_{N}^y\bigr]  + J^z   \sigma_{N-1}^z   \sigma_{N}^z      \nonumber\\
\fl &H_{{\rm  media}} = \sum_{i=2}^{N-2} ( J_i^x \sigma_i^x   \sigma_{i+1}^x + J_i^y \sigma_i^y   \sigma_{i+1}^y+ J_i^z \sigma_i^z   \sigma_{i+1}^z) + \sum_{i=2}^{N-1} h_i^z \sigma_i^z ,     \label{Definition_H_media_couple}
\end{eqnarray} 
where $J\equiv J^x+J^y$ and  $\gamma\equiv\frac{J^x-J^y}{J^x+J^y}$.
Now, we consider the term $H_{{\rm  couple}}$, which couples the focused spins and the mediator spins, as perturbation and carry out perturbation expansion for the eigenvalues and the eigenstates~\cite{Density_matrix_expansion}.
The unperturbed density matrix $\rho_{{\rm  tot}}^{(0)}$ is given by
\begin{eqnarray}
\rho_{{\rm  tot}}^{(0)} =  e^{-\beta H_1-\beta H_N}  e^{-\beta H_{{\rm  media}}} \label{unperturbed_denstity_matrix_case_b}
\end{eqnarray} 
because $H_1$, $H_{{\rm  media}}$ and $H_N$ commute with each other.

Because the external fields are applied in the \textit{z} direction, the unperturbed eigenstates of $H_1+ H_N$ are given by ${\{\ket{{\ua_1\ua_N}},\ket{{\ua_1\da_N}} ,\ket{{\da_1\ua_N}},\ket{{\da_1\da_N}}\}}$ with the corresponding eigenvalues $\{-h_{1\op}-h_{N\op},-h_{1\op}+h_{N\op},h_{1\op}-h_{N\op},h_{1\op}+h_{N\op}\}$; we denote these unperturbed eigenvalues as $\{E_{1N}^{\ua\ua},E_{1N}^{\ua\da},E_{1N}^{\da\ua},E_{1N}^{\da\da}\}$. 
We also define the unperturbed eigenstates of $H_{{\rm  media}}$ as 
\begin{eqnarray}
\ket{\psi_{{\rm  media}}^n}&= s_n \ket{{\ua_2}} \ket{\tilde{\psi}_n^{\ua\ua}} \ket{{\ua_{N-1}}}  + t_n \ket{{\ua_2}} \ket{\tilde{\psi}_n^{\ua\da}} \ket{{\da_{N-1}}}\nonumber \\
 &+ u_n \ket{{\da_2}} \ket{\tilde{\psi}_n^{\da\ua}} \ket{{\ua_{N-1}}} + w_n \ket{{\da_2}} \ket{\tilde{\psi}_n^{\da\da}} \ket{{\da_{N-1}}},    \label{Eigenstates_of_media_spin_system}
\end{eqnarray} 
for $n=1,2,\cdots,2^{N-2}$, where $\{\ket{\tilde{\psi}_n^{\ua\ua}},\ket{\tilde{\psi}_n^{\ua\da}},\ket{\tilde{\psi}_n^{\da\ua}},\ket{\tilde{\psi}_n^{\da\da}}\}$ are the states of the spins from $3$ to $N-2$.
Because the total Hamiltonian $H_{{\rm  tot}}$ is a real matrix, the coefficients $\{s_n,t_n,u_n,w_n\}$ in Eq.~\eqref{Eigenstates_of_media_spin_system} are real numbers.
We define the corresponding unperturbed eigenvalues of $H_{{\rm  media}}$ as $\{E_\md^n\}$.
Then, the unperturbed eigenstates of the total system are given by 
\begin{eqnarray}
\fl \{\ket{{\ua_1\ua_N}}\otimes\ket{\psi_{{\rm  media}}^n}, \ket{{\ua_1\da_N}}\otimes\ket{\psi_{{\rm  media}}^n}, \ket{{\da_1\ua_N}}\otimes\ket{\psi_{{\rm  media}}^n}, \ket{{\da_1\da_N}}\otimes\ket{\psi_{{\rm  media}}^n}\}  \label{unperturbed_states_case_b}
\end{eqnarray}
with the unperturbed eigenvalues, $E_{1N}^{\ua\ua}+E_\md^n$ $etc$.

We then define the perturbed eigenstates corresponding to each of \eqref{unperturbed_states_case_b} as $\{\ket{\psi_{{\rm  tot}}^{n,\ua\ua}}, \ket{\psi_{{\rm  tot}}^{n,\ua\da}},\ket{\psi_{{\rm  tot}}^{n,\da\ua}}, \ket{\psi_{{\rm  tot}}^{n,\da\da}}\}$, respectively. 
We express them as
\begin{eqnarray}
\ket{\psi_{{\rm  tot}}^{n,\xi}} &= \ket{{\ua_1\ua_N}}\otimes \ket{\psi_{{\rm  media},\ua\ua}^{n,\xi}}+\ket{{\ua_1\da_N}}\otimes \ket{\psi_{{\rm  media},\ua\da}^{n,\xi}}\nonumber \\
    &+ \ket{{\da_1\ua_N}}\otimes \ket{\psi_{{\rm  media},\da\ua}^{n,\xi}} + \ket{{\da_1\da_N}}\otimes \ket{\psi_{{\rm  media},\da\da}^{n,\xi}} ,  \label{Eigenstates_of_total_spin_system}
\end{eqnarray} 
for $n=1,2,\cdots,2^{N-2}$ and $\xi=\ua\ua,\ua\da,\da\ua,\da\da$,
where $\ket{\psi_{{\rm  media},\ua\ua}^{n,\xi}}$, $\ket{\psi_{{\rm  media},\ua\da}^{n,\xi}}$, $\ket{\psi_{{\rm  media},\da\ua}^{n,\xi}}$ and $\ket{\psi_{{\rm  media},\da\da}^{n,\xi}}$ are the states of the spins from $2$ to $N-1$ and may not be normalized.
We also define the corresponding perturbed eigenvalues as $\{E_{{\rm  tot}}^{n,\ua\ua},E_{{\rm  tot}}^{n,\ua\da},E_{{\rm  tot}}^{n,\da\ua},E_{{\rm  tot}}^{n,\da\da}\}$, which we express as
\begin{eqnarray}
E_{{\rm  tot}}^{n,\xi} = E_{1N}^\xi+E_\md^n +\delta E_{{\rm  tot}}^{n,\xi} \label{Eigenvalues_of_total_spin_system}
\end{eqnarray}
for $n=1,2,\cdots,2^{N-2}$ and $\xi=\ua\ua,\ua\da,\da\ua,\da\da$.
Note that $\{E_\md^n\}$ and $\{\delta E_{{\rm  tot}}^{n,\xi}\}$ do not depend on the temperature $\beta$ because $H_{{\rm  media}}$ and $H_{{\rm  couple}}$ do not depend on $\beta$.
Then, we can say that 
\begin{eqnarray}
 \frac{  e^{ -\beta E_{{\rm  tot}}^{n,\ua\ua} }  }{Z_{{\rm  tot}} }  =  \frac{  e^{ \beta (h_{1\op}+h_{N\op}) - \beta(E_\md^n +\delta E_{{\rm  tot}}^{n,\ua\ua}) }  }{Z_{{\rm  tot}} }\stackrel{\beta\to 0}{\longrightarrow}{\rm const} . \label{order_of_max_mixedness}
\end{eqnarray}
in the limit $\beta\rightarrow0$ in the case $h_{1\op}, h_{N\op}>0$, where $Z_{{\rm  tot}}$ is the partition function of the total Hamiltonian.

Now, we show the outline of the proof.
First, we calculate the elements $\{F_1,F_2, p_{\ua\ua},p_{\ua\da}, p_{\da\ua},p_{\da\da} \}$ from Eqs.~\eqref{Eigenstates_of_total_spin_system} and \eqref{Eigenvalues_of_total_spin_system} as
\begin{eqnarray}
\fl F_1= \sum_{n=1}^{2^{N-2}} \sum_{\xi=\ua\ua,\ua\da,\da\ua,\da\da} \frac{e^{-\beta E_{{\rm  tot}}^{n,\xi}}}{Z_\tot} F_1^{n,\xi}  \quad
{\rm and} \quad
p_{\ua\ua}=\sum_{n=1}^{2^{N-2}} \sum_{\xi=\ua\ua,\ua\da,\da\ua,\da\da}  \frac{e^{-\beta E_{{\rm  tot}}^{n,\xi}}}{Z_\tot}   p_{\ua\ua}^{n,\xi},  \label{Cal_F_1_and_p}
\end{eqnarray}
for example, where each contribution is given by
\begin{eqnarray}
\fl F_1^{n,\xi}= \bigl \langle \psi_{{\rm  media},\ua\da}^{n,\xi} \bigl | \psi_{{\rm  media},\da\ua}^{n,\xi} \bigr \rangle \quad {\rm and} \quad p_{\ua\ua}^{n,\xi}= \bigl \langle \psi_{{\rm  media},\ua\ua}^{n,\xi} \bigl | \psi_{{\rm  media},\ua\ua}^{n,\xi} \bigr \rangle. \label{Cal_F_1_and_p_cont}
\end{eqnarray} 
We define $\{F_2^{n,\xi}, p_{\ua\da}^{n,\xi}, p_{\da\ua}^{n,\xi},p_{\da\da}^{n,\xi} \}$ in the same way.
Since the elements $\{F_1,F_2, p_{\ua\ua},p_{\ua\da}, p_{\da\ua},p_{\da\da} \}$ are additive with respect to the indices $n$ and $\xi$, we calculate each contribution in Eq.~\eqref{Cal_F_1_and_p_cont} separately.

Second, we estimate the leading orders of $\{F_1,F_2\}$ and $\{p_{\ua\ua},p_{\ua\da}, p_{\da\ua},p_{\da\da}\}$ with respect to $\beta$.
From the perturbation theory, we obtain the approximate forms of $\{\ket{\psi_{{\rm  tot}}^{n,\xi}}\}$ and expand $\{F_1,F_2\}$ and $\{p_{\ua\ua},p_{\ua\da}, p_{\da\ua},p_{\da\da} \}$ with respect to $\beta$.
We thus prove the statements \eqref{Paraorders1} and \eqref{Paraorders2}.
For this purpose, we first calculate $\ket{\psi_{{\rm  tot}}^{n_0,\ua\ua}}$ in Eq.~\eqref{Eigenstates_of_total_spin_system}, namely the perturbed state of $\ket{{\ua_1\ua_N}}\otimes \ket{\psi_{{\rm  media}}^{n_0}}$.
Then, we obtain the contributions of $\ket{\psi_{{\rm  tot}}^{n_0,\ua\ua}}$ to the elements $\{F_1,F_2\}$ and $\{p_{\ua\ua},p_{\ua\da},p_{\da\ua},p_{\da\da}\}$, which are defined as $\{F_1^{n_0,\ua\ua},F_2^{n_0,\ua\ua}\}$ and $\{p_{\ua\ua}^{n_0,\ua\ua},p_{\ua\da}^{n_0,\ua\ua},p_{\da\ua}^{n_0,\ua\ua},p_{\da\da}^{n_0,\ua\ua}\}$ in Eq.~\eqref{Cal_F_1_and_p_cont}. 
From the calculation in Sec.~1 of the supplementary materials, we obtain the leading terms of the elements $\{F_1,F_2\}$ and $\{p_{\ua\ua},p_{\ua\da},p_{\da\ua},p_{\da\da}\}$ as
\begin{eqnarray}
\fl F_1^{n_0,\ua\ua} =& \frac{J^2}{4h_{1\op} h_{N\op}}  \Bigl( \gamma s_{n_0} w_{n_0} \langle \tilde{\psi}_{n_0}^{\da\da} | \tilde{\psi}_{n_0}^{\ua\ua} \rangle  +u_{n_0}t_{n_0} \langle   \tilde{\psi}_{n_0}^{\ua\da} | \tilde{\psi}_{n_0}^{\da\ua} \rangle   \nonumber \\
\fl&+ \gamma^2 t_{n_0} u_{n_0} \langle   \tilde{\psi}_{n_0}^{\da\ua}  |\tilde{\psi}_{n_0}^{\ua\da} \rangle+  \gamma w_{n_0} s_{n_0} \langle   \tilde{\psi}_{n_0}^{\ua\ua}  |\tilde{\psi}_{n_0}^{\da\da} \rangle + O(\beta^{\kappa_N}) \Bigr)  ,      \label{Approximate_form_of_F1_n0} \\
\fl F_2^{n_0,\ua\ua} =& \frac{J^2}{4h_{1\op} h_{N\op}}  \Bigl(\gamma^2 w_{n_0}  s_{n_0} \langle \tilde{\psi}_{n_0}^{\da\da} | \tilde{\psi}_{n_0}^{\ua\ua} \rangle+\gamma u_{n_0} t_{n_0}   \langle \tilde{\psi}_{n_0}^{\da\ua} | \tilde{\psi}_{n_0}^{\ua\da} \rangle    \nonumber\\
 \fl&+\gamma t_{n_0} u_{n_0}   \langle \tilde{\psi}_{n_0}^{\ua\da} | \tilde{\psi}_{n_0}^{\da\ua} \rangle +  s_{n_0}  w_{n_0} \langle \tilde{\psi}_{n_0}^{\ua\ua} | \tilde{\psi}_{n_0}^{\da\da} \rangle  + O(\beta^{\kappa_N})  \Bigr)   \label{Approximate_form_of_F2_n0}
\end{eqnarray}
as well as the elements $\{p_{\ua\ua}^{n_0,\ua\ua},p_{\ua\da}^{n_0,\ua\ua},p_{\da\ua}^{n_0,\ua\ua},p_{\da\da}^{n_0,\ua\ua}\}$ in the forms  
\begin{eqnarray}
\fl&p_{\ua\ua}^{n_0,\ua\ua}=1+O(\beta^{2\kappa_N}) ,\quad p_{\ua\da}^{n_0,\ua\ua}=\frac{J^2}{4h_{N\op}^2 } \Bigl(\gamma^2  s_{n_0}^2 + t_{n_0}^2 + \gamma^2 u_{n_0}^2+  w_{n_0}^2 +O\bigl(\beta^{\kappa_N}\bigr)  \Bigr) ,\nonumber \\
\fl& p_{\da\ua}^{n_0,\ua\ua}=\frac{J^2}{4h_{1\op}^2 } \Bigl(\gamma^2  s_{n_0}^2 +\gamma^2  t_{n_0}^2 + u_{n_0}^2 +  w_{n_0}^2 +O(\beta^{\kappa_1})+O\bigl(\beta^{2\kappa_N}\bigr)   \Bigr),\nonumber \\
\fl& p_{\da\da}^{n_0,\ua\ua}=\frac{J^4}{16h_{1\op}^2 h_{N\op}^2}   \Bigl(\gamma^4 s_{n_0}^2 + \gamma^2 t_{n_0}^2 +\gamma^2 u_{n_0}^2 + w_{n_0}^2 +O(\beta^{\kappa_N})     \Bigr).   \label{inequality_for_p_uada_daua_case2}
\end{eqnarray}
We utilized the notation $O$ in the following sense; if $f(\beta) = O(\beta^x)$, the function $f(\beta)$ is of order of $\beta^x$ or higher.

Then, we sum the elements $\{F_1^{n,\ua\ua},F_2^{n,\ua\ua}\}$ and $\{p_{\ua\ua}^{n,\ua\ua},p_{\ua\da}^{n,\ua\ua},p_{\da\ua}^{n,\ua\ua},p_{\da\da}^{n,\ua\ua}\}$ with the Boltzmann weight $e^{-\beta E_{{\rm  tot}}^{n,\ua\ua}}$ over the label $n$ accordingly to Eq~\eqref{Cal_F_1_and_p}.
First, we calculate the summation of each of $\{F_1^{n,\ua\ua},F_2^{n,\ua\ua}\}$.
Because the spins~2 and ($N-1$) are separated by ($N-4$) spins, the correlation between the spins~2 and ($N-1$) are generated by the ($N-3$)th-order perturbation of $H_{{\rm  media}}$.
Therefore, we obtain
\begin{eqnarray}
\langle \sigma_2^x  \sigma_{N-1}^x \rangle_0  = O(\beta ^{\alpha_1}), \quad \langle \sigma_2^y  \sigma_{N-1}^y \rangle_0  = O(\beta ^{\alpha_2}),     
\end{eqnarray}
where $\langle \cdots \rangle_0$ denotes the thermal average with respect to $\rho_{{\rm  tot}}^{(0)}$ in \eqref{unperturbed_denstity_matrix_case_b} and $\alpha_1 \ge N-3$, $\alpha_2 \ge N-3$. 
Since we are considering the case $N\ge4$, we have $\alpha_1\ge1$ and $\alpha_2\ge1$.
This is the key to the fact that the maximized entanglement vanishes for $N\ge4$. 

From the equations
\begin{eqnarray}
\fl\frac{\langle \sigma_2^x  \sigma_{N-1}^x  + \sigma_2^y  \sigma_{N-1}^y \rangle_0}{4}&=\tr \biggl(  e^{-\beta H_{{\rm  media}}} \frac{\langle \sigma_2^+  \sigma_{N-1}^-  + \sigma_2^-  \sigma_{N-1}^+ \rangle_0}{2}\biggr)\nonumber\\
\fl&= \sum_{n=1}^{2^{N-2}} \frac{e^{-\beta E_{{\rm  media}}^n}}{Z_\md}  \bigl \langle \ua_2\da_{N-1} \bigl | \psi_{{\rm  media}}^{n}  \bigr \rangle \bigl \langle  \psi_{{\rm  media}}^{n}  \bigl |  \da_2\ua_{N-1}   \bigr \rangle,       \label{Relation_between_correlation_F1}  \\
\fl \frac{\langle \sigma_2^x  \sigma_{N-1}^x  - \sigma_2^y  \sigma_{N-1}^y \rangle_0}{4}&=\sum_{n=1}^{2^{N-2}} \frac{e^{-\beta E_{{\rm  media}}^n}}{Z_\md}   \bigl \langle \da_2\da_{N-1} \bigl | \psi_{{\rm  media}}^{n}  \bigr \rangle \bigl \langle  \psi_{{\rm  media}}^{n}  \bigl |  \ua_2\ua_{N-1}   \bigr \rangle,\label{Relation_between_correlation_F2}
\end{eqnarray}
where $Z_\md\equiv \tr (e^{-\beta H_{{\rm  media}}})$, $\sigma^{+}\equiv (\sigma^x+i \sigma^y)/2$ and $\sigma^{-}\equiv (\sigma^x-i \sigma^y)/2$, 
we also have
\begin{eqnarray}
 \fl\sum_{n=1}^{2^{N-2}} e^{-\beta E_\md^n} t_n u_n  \langle \tilde{\psi}_n^{\ua\da} | \tilde{\psi}_n^{\da\ua} \rangle  = O(\beta^{\alpha}), \quad 
\sum_{n=1}^{2^{N-2}} e^{-\beta E_\md^n} s_n w_n  \langle \tilde{\psi}_n^{\da\da} | \tilde{\psi}_n^{\ua\ua} \rangle = O(\beta^{\alpha}) , \label{Oalpha1}
\end{eqnarray}
where $\alpha=\min(\alpha_1,\alpha_2)$.
Moreover, because $\{\delta E_{{\rm  tot}}^{n,\xi}\}$ in Eq.~\eqref{Eigenvalues_of_total_spin_system} do not depend on $\beta$, we have
\begin{eqnarray}
e^{-\beta E_{{\rm  tot}}^{n,\xi}} = e^{-\beta (E_{1N}^{\xi} + E_{\md}^n)} \Bigl(1-\beta \delta E_{{\rm  tot}}^{n,\xi} + O(\beta^2) \Bigr) ,\label{restriction_weight}
\end{eqnarray}
for $n=1,2,\cdots,2^{N-2}$ and $\xi=\ua\ua,\ua\da,\da\ua,\da\da$.
Then, we obtain from Eqs.~\eqref{Approximate_form_of_F1_n0}, \eqref{Approximate_form_of_F2_n0}, \eqref{Oalpha1} and \eqref{restriction_weight},
\begin{eqnarray}
\fl&\sum_{n=1}^{2^{N-2}}  e^{-\beta E_{{\rm  tot}}^{n,\ua\ua}} F_1^{n,\ua\ua} = \sum_{n=1}^{2^{N-2}}   e^{-\beta (E_{1N}^{\ua\ua} + E_{\md}^n)}  \bigl(1-\beta \delta E_{{\rm  tot}}^{n,\ua\ua} + O(\beta^2) \bigr)F_1^{n,\ua\ua}   \nonumber\\
\fl=&e^{-\beta E_{1N}^{\ua\ua}} \frac{J^2}{4h_{1\op} h_{N\op}}  \Bigl( O(\beta^{\alpha}) + O(\beta^{\kappa_N}) +  O(\beta) \Bigr)=e^{-\beta E_{1N}^{\ua\ua}}\times O(\beta^{\kappa_1+\kappa_N+\kappa})     \label{Final_form_uaua_F1_case_b}
\end{eqnarray}
and 
\begin{eqnarray}
\sum_{n=1}^{2^{N-2}} e^{-\beta E_{{\rm  tot}}^{n,\ua\ua}} F_2^{n,\ua\ua} =e^{-\beta E_{1N}^{\ua\ua}}\times  O(\beta^{\kappa_1+\kappa_N+\kappa}),    \label{Final_form_uaua_F2_case_b}
\end{eqnarray}
where $\alpha=\min(\alpha_1,\alpha_2)\ge1 $ and $\kappa$ is defined in \eqref{Definition_of_the_two_kappa}.

We similarly calculate the contributions of the other states $\{\ket{\psi_{{\rm  tot}}^{n,\xi}}\}$ (${\xi=\ua\da,\da\ua,\da\da}$).
From the calculation in Sec.~1 of the supplementary materials, we obtain the leading terms of the elements $\{\ket{\psi_{{\rm  media},\xi}^{n_0,\ua\da}}\}$ for ${\xi=\ua\ua, \ua\da,\da\ua,\da\da}$ in Eq.~\eqref{Eigenstates_of_total_spin_system}, and have the contributions of the states $\{\ket{\psi_{{\rm  tot}}^{n,\ua\da}}\}$ to $F_1$ and $F_2$ as
 \begin{eqnarray}
\fl \sum_{n=1}^{2^{N-2}}  e^{-\beta E_{{\rm  tot}}^{n,\ua\da}} F_1^{n,\ua\da}&= e^{-\beta E_{1N}^{\ua\da}} \frac{J^2}{4h_{1\op} h_{N\op}}  \Bigl( O(\beta^{\alpha}) + O(\beta^{\kappa_N}) + O(\beta^{\tilde{\kappa}}) + O(\beta)   \Bigr)\nonumber \\
\fl &=e^{-\beta E_{1N}^{\ua\da}}\times O(\beta^{\kappa_1+\kappa_N+\kappa'})  , \nonumber\\
\fl\sum_{n=1}^{2^{N-2}}  e^{-\beta E_{{\rm  tot}}^{n,\ua\da}} F_2^{n,\ua\da} &=e^{-\beta E_{1N}^{\ua\da}} \frac{J^2}{4h_{1\op} h_{N\op}}  \Bigl( O(\beta^{\alpha}) + O(\beta^{\kappa_N}) + O(\beta)  \Bigr)\nonumber \\
\fl&=e^{-\beta E_{1N}^{\ua\da}}  \times O(\beta^{\kappa_1+\kappa_N+\kappa}) , \label{Final_form_daua_F1F2_case_b}
\end{eqnarray}
where we utilized Eqs.~\eqref{Oalpha1} and \eqref{restriction_weight}, and $\kappa'$ is defined in \eqref{Definition_of_the_two_kappa}.
From the inequality
$
e^{-\beta E_{1N}^{\xi}}/Z_\tot<1
$
for $\xi=\ua\ua,\ua\da,\da\ua,\da\da$, we finally obtain \eqref{Paraorders1} by substituting Eqs.~\eqref{Final_form_uaua_F1_case_b}--\eqref{Final_form_daua_F1F2_case_b} into Eq.~\eqref{Cal_F_1_and_p}.

Second, we calculate the summation of each of $\{p_{\ua\ua}^{n,\ua\ua},p_{\ua\da}^{n,\ua\ua},p_{\da\ua}^{n,\ua\ua},p_{\da\da}^{n,\ua\ua}\}$. 
From Eqs.~\eqref{order_of_max_mixedness}, \eqref{Cal_F_1_and_p} and \eqref{inequality_for_p_uada_daua_case2}, we obtain 
\begin{eqnarray}
\fl p_{\ua\ua} \ge \sum_{n=1}^{2^{N-1}}  \frac{e^{-\beta E_{{\rm  tot}}^{n,\ua\ua}}}{Z_{{\rm  tot}}}  p_{\ua\ua}^{n,\ua\ua} =  \sum_{n=1}^{2^{N-1}}  \frac{e^{-\beta E_{{\rm  tot}}^{n,\ua\ua}}}{Z_{{\rm  tot}}}   \Bigl[1 +O\bigl(\beta^{2\kappa_N}\bigr)  \Bigr] \ge W_{\ua\ua}  , 
\end{eqnarray}
where we define a positive number independent of $\beta$ as $W_{\ua\ua}$ so as to satisfy the above inequalities in the limit of $\beta \to0$.
Next, because the quantity $\gamma^2  s_{n}^2 +\gamma^2  t_{n}^2 + u_{n}^2 +  w_{n}^2$ cannot vanish for all $n$, we have
\begin{eqnarray}
\fl p_{\ua\da} &\ge \sum_{n=1}^{2^{N-1}}  \frac{e^{-\beta E_{{\rm  tot}}^{n,\ua\ua}}}{Z_{{\rm  tot}}}  p_{\ua\ua}^{n,\ua\da}  \nonumber \\
\fl &   =   \frac{J^2}{4h_{1\op}^2 }  \sum_{n=1}^{2^{N-1}}  \frac{e^{-\beta E_{{\rm  tot}}^{n,\ua\ua}}}{Z_{{\rm  tot}}} \Bigl(\gamma^2  s_{n}^2 +\gamma^2  t_{n}^2 + u_{n}^2 +  w_{n}^2 +O(\beta^{\kappa_N})  \Bigr) \ge  \frac{J^2}{4h_{1\op}^2 }   W_{\ua\da}  , 
\end{eqnarray}
where we define another positive number independent of $\beta$ as $W_{\ua\da}$ so as to satisfy the above inequalities in the limit of $\beta \to0$.
Similarly, we can obtain
\begin{eqnarray}
p_{\da\ua}\ge  \frac{J^2}{4h_{N\op}^2 }   W_{\da\ua}      , \quad   p_{\da\da}\ge  \frac{J^4}{16h_{1\op}^2 h_{N\op}^2 }   W_{\da\da}     .     
\end{eqnarray}
 where $W_{\da\ua}$ and $W_{\da\da}$ are positive numbers which do not depend on $\beta$.
From the above inequality, we have
\begin{eqnarray}
p_{\ua\da} p_{\da\ua} \ge  \frac{J^4}{16h_{1\op}^2 h_{N\op}^2 }   W_{\ua\da}   W_{\da\ua}      , \quad   p_{\ua\ua} p_{\da\da} \ge  \frac{J^4}{16h_{1\op}^2 h_{N\op}^2 }   W_{\ua\ua}  W_{\da\da}     .     \label{inequality_for_P} 
\end{eqnarray}
 in the limit $\beta\rightarrow0$.
If $p_{\ua\da} p_{\da\ua}$ is of order higher than $\beta^{2\kappa_1+2\kappa_N}$, $p_{\ua\da} p_{\da\ua}$ decays faster than $J^4W_{\ua\da}   W_{\da\ua}/(16h_{1\op}^2 h_{N\op}^2)$, and hence, the inequality \eqref{inequality_for_P} is not satisfied in the limit $\beta\rightarrow0$.
Therefore, $p_{\ua\da} p_{\da\ua}$ must be of order of $\beta^{2\kappa_1+2\kappa_N}$ or \textit{lower}.
We can similarly prove that $p_{\ua\ua} p_{\da\da}$ is of order of $\beta^{2\kappa_1+2\kappa_N}$ or \textit{lower}.
Thus, we obtain \eqref{Paraorders2}.

We thereby prove from~\eqref{Paraorders1} and \eqref{Paraorders2} that the inequality~\eqref{Prove_this_inequality} is satisfied below a certain value of $\beta$ in the case~(b).

%%%%%%%%%%%%%%%%%%%%%%%%%%%%%%%%%%%%%%%%%%%%%%%%%%%%%%%%%%%%%%%%%%%%%%%%%%%%%%%%%%%%%%%%%%%%%%%%%%%%%%%%%%%%%%%%
%%%%%%%%%%%%%%%%%%%%%%%%%%%%%%%%%%%%%%%%%%%%%%%%%%%%%%%%%%%%%%%%%%%%%%%%%%%%%%%%%%%%%%%%%%%%%%%%%%%%%%%%%%%%%%%%

\textit{Case~(c)}. 
To simplify the problem, we consider the case of $h_1>0$, $\kappa_1\ge1$ and $\kappa_N\le0$, but we can prove the other cases in the same way.
In this case, we prove the following; $F_{1}$ and $F_{2}$ are both of order of
\begin{eqnarray}
\beta^{\kappa_1+1}  \  {\rm or} \ {\rm  higher}.   \label{Paraorders21}
\end{eqnarray}
On the other hand, $p_{\ua\ua}p_{\da\da}$ and $p_{\ua\da}p_{\da\ua}$ are both of order of
\begin{eqnarray}
\beta^{2\kappa_1} \  {\rm or} \ {\rm  lower}.     \label{Paraorders22}
\end{eqnarray} 
Then, the inequality~\eqref{Prove_this_inequality} is satisfied below a certain value of $\beta$.

In a similar manner to the case~(b), we separate the total Hamiltonian as follows:
\begin{eqnarray}
H_{{\rm  tot}} = H_1 +H'_{{\rm  couple}} +  H'_{{\rm  media}},
\end{eqnarray} 
where 
\begin{eqnarray}
\fl&H_1= h_{1\op} \sigma_1^z,  \quad   H'_{{\rm  couple}} = J^x \sigma_1^x   \sigma_{2}^x + J^y \sigma_1^y   \sigma_{2}^y + J^z   \sigma_1^z   \sigma_{2}^z , \nonumber\\
\fl&H'_{{\rm  media}} = \sum_{i=2}^{N-1} ( J_i^x \sigma_i^x   \sigma_{i+1}^x + J_i^y \sigma_i^y   \sigma_{i+1}^y+ J_i^z \sigma_i^z   \sigma_{i+1}^z)   + \sum_{i=2}^{N-1} h_i^z \sigma_i^z+h_{N\op} \sigma_N^z .  
\end{eqnarray} 
Note that in the case~(c), the norm of the Hamiltonian $H'_{{\rm  media}}$ is of order of $\beta^0$ because $\kappa_N\le0$.

Here, we consider the interaction term $H'_{{\rm  couple}}$, which couples the spins 1 and the other spins, as perturbation.
Then, the unperturbed density matrix $\rho_{{\rm  tot}}^{'(0)}$ is expressed as
$
\rho_{{\rm  tot}}^{'(0)} =  e^{-\beta H_1}  e^{-\beta H'_{{\rm  media}}} .
$
The unperturbed eigenstates of $H_1$ are given by ${\{\ket{{\ua_1}},\ket{{\da_1}}\}}$ with the corresponding unperturbed eigenvalues $\{-h_{1\op},h_{1\op}\}$; we denote these eigenvalues as $\{E_1^{\ua},E_1^{\da}\}$.
We also define the unperturbed eigenstates of $H'_{{\rm  media}}$ as
\begin{eqnarray}
\fl\ket{\phi_\md^n} = s'_n\ket{{\ua_2}} \ket{\tilde{\phi}_n^{\ua\ua}} \ket{{\ua_N}}  + t'_n \ket{{\ua_2}} \ket{\tilde{\phi}_n^{\ua\da}} \ket{{\da_N}} + u'_n \ket{{\da_2}} \ket{\tilde{\phi}_n^{\da\ua}} \ket{{\ua_N}} + w'_n \ket{{\da_2}} \ket{\tilde{\phi}_n^{\da\da}} \ket{{\da_N}} \label{definition_of_Phi_proof2}
\end{eqnarray} 
for $n=1,2,\cdots2^{N-1}$, where $\{ \ket{\tilde{\phi}_n^{\ua\ua}}, \ket{\tilde{\phi}_n^{\ua\da}}, \ket{\tilde{\phi}_n^{\da\ua}}, \ket{\tilde{\phi}_n^{\da\da}}\}$ are the states of the spins from $3$ to $N-1$.
Because the total Hamiltonian $H_{{\rm  tot}}$ is a real matrix, the coefficients $\{s'_n,t'_n,u'_n,w'_n\}$ are real numbers.
We define the unperturbed eigenvalues of $H'_{{\rm  media}}$ as $\{E_\md^{'n}\}$.
Then, the unperturbed eigenstates of the total system are given by $\{\ket{{\ua_1}}\otimes\ket{\phi_{{\rm  media}}^n}\}$ and $\{\ket{{\da_1}}\otimes\ket{\phi_{{\rm  media}}^n}\}$.

We then define the corresponding perturbed eigenstates as $\{\ket{\phi_{{\rm  tot}}^{n,\ua}}\}$ and $\{\ket{\phi_{{\rm  tot}}^{n,\da}}\}$ and the corresponding perturbed eigenvalues as $\{E_{{\rm  tot}}^{n,\ua},E_{{\rm  tot}}^{n,\da}\}$. We express them in the forms 
\begin{eqnarray}
\ket{\phi_{{\rm  tot}}^{n,\eta}} =\ket{{\ua_1}} \otimes \ket{\phi_{{\rm  media},\ua}^{n,\eta}} + \ket{{\da_1}} \otimes \ket{\phi_{{\rm  media},\da}^{n,\eta}}     \label{Eigenstates_of_total_spin_system2}
\end{eqnarray}
and 
\begin{eqnarray}
E_{{\rm  tot}}^{'n,\eta} = E_1^\eta+E_\md^{'n} +\delta E_{{\rm  tot}}^{'n,\eta} \label{Eigenvalues_of_total_spin_system2}
\end{eqnarray}
for $n=1,2,\cdots2^{N-1}$ and $\eta=\ua,\da$, where $\ket{\phi_{{\rm  media},\ua}^{n,\eta}}$ and $\ket{\phi_{{\rm  media},\da}^{n,\eta}}$ are the states of the spins from 2 to $N$ and may  not be normalized.
Note that $\{E_\md^n\}$ and $\{\delta E_{{\rm  tot}}^{'n,\xi}\}$ are of order of $\beta^0$ because $||H'_{{\rm  media}}||$ and $||H'_{{\rm  couple}}||$ are of order of $\beta^0$.
Then, in the limit $\beta\rightarrow0$,  we can say that 
\begin{eqnarray}
\frac{  e^{ -\beta E_{{\rm  tot}}^{'n,\ua} }  }{Z_\tot}  =  \frac{  e^{ \beta h_{1\op} - \beta(E_\md^{'n} +\delta E_{{\rm  tot}}^{'n,\ua}) }  }{Z_\tot}   \stackrel{\beta\to 0}{\longrightarrow}{\rm const}   \label{order_of_max_mixedness2}
\end{eqnarray}
in the case $h_{1\op}>0$.

Next, we calculate the elements $\{F_1,F_2\}$ and $\{p_{\ua\ua},p_{\ua\da},p_{\da\ua},p_{\da\da}\}$.
As in the case~(b), they are additive with respect to the indices $n$ and $\eta$, and hence we define each contribution of $\ket{\phi_{{\rm  tot}}^{n,\eta}}$ to the elements $\{F_1,F_2\}$ and $\{p_{\ua\ua},p_{\ua\da},p_{\da\ua},p_{\da\da}\}$ as $\{F_1^{n,\eta},F_2^{n,\eta}\}$ and $\{p_{\ua\ua}^{n,\eta},p_{\ua\da}^{n,\eta},p_{\da\ua}^{n,\eta},p_{\da\da}^{n,\eta}\}$.
The elements $\{F_1,F_2\}$ and $\{p_{\ua\ua},p_{\ua\da},p_{\da\ua},p_{\da\da}\}$ are given by
\begin{eqnarray}
\fl F_1=\sum_{n=1}^{2^{N-1}}\sum_{\eta=\ua,\da} \frac{  e^{ -\beta E_{{\rm  tot}}^{'n,\ua} }  }{Z_\tot}  F_1^{n,\eta} \quad
{\rm and} 
\quad 
p_{\ua\ua}=\sum_{n=1}^{2^{N-1}} \sum_{\eta=\ua,\da}  \frac{  e^{ -\beta E_{{\rm  tot}}^{'n,\ua} }  }{Z_\tot}  p_{\ua\ua}^{n,\eta},       \label{additive_P_case_c}
\end{eqnarray} 
for example. Each contribution is given by
\begin{eqnarray}
F_1^{n,\eta} = \tr \Bigl( \ket{{\da_1\ua_N}} \bra{{\ua_1\da_N}}\otimes I_\md  \ket{\phi_{{\rm  tot}}^{n,\eta}}\bra{\phi_{{\rm  tot}}^{n,\eta}} \Bigr)
\end{eqnarray} 
and 
\begin{eqnarray}
p_{\ua\ua}^{n,\eta}=\tr \Bigl( \ket{{\ua_1\ua_N}} \bra{{\ua_1\ua_N}}\otimes I_\md  \ket{\phi_{{\rm  tot}}^{n,\eta}}\bra{\phi_{{\rm  tot}}^{n,\eta}} \Bigr) ,
\end{eqnarray} 
where $I_\md$ is the identity operator in the whole space of the mediator spins.

Now, we calculate each contribution of $\ket{\phi_{{\rm  tot}}^{n_0,\ua}}$ to the elements $\{F_1,F_2\}$ and $\{p_{\ua\ua},p_{\ua\da},p_{\da\ua},p_{\da\da}\}$, which is given by $\{F_1^{n_0,\ua},F_2^{n_0,\ua}\}$ and $\{p_{\ua\ua}^{n_0,\ua},p_{\ua\da}^{n_0,\ua},p_{\da\ua}^{n_0,\ua},p_{\da\da}^{n_0,\ua}\}$.
The calculation in Sec.~2 of the supplementary materials gives them as
\begin{eqnarray}   
\fl F_1^{n_0,\ua} =&\frac{J}{-2h_{1\op} } \Bigl( \gamma s'_{n_0} w_{n_0}' \langle \phi_{n_0}^{\da\da} | \phi_{n_0}^{\ua\ua} \rangle + u'_{n_0}t_{n_0}' \langle   \phi_{n_0}^{\da\ua} | \phi_{n_0}^{\ua\da} \rangle + O(\beta^{\kappa_1}) \Bigr) ,      \nonumber \\
\fl F_2^{n_0,\ua} =&\frac{J}{-2h_{1\op} }  \Bigl( \gamma t'_{n_0} u_{n_0}' \langle \phi_{n_0}^{\da\ua} | \phi_{n_0}^{\ua\da} \rangle  + w'_{n_0}s_{n_0}' \langle   \phi_{n_0}^{\ua\ua} | \phi_{n_0}^{\da\da} \rangle + O(\beta^{\kappa_1})  \Bigr)   \label{perturbed_form_of_F2_n0_ua}
\end{eqnarray}
and 
\begin{eqnarray}   
\fl p_{\ua\ua}^{n_0,\ua}=&s'^{2}_{n_0}+u'^{2}_{n_0} +O(\beta^{2\kappa_1}),   \quad p_{\ua\da}^{n_0,\ua}=t'^{2}_{n_0}+w'^{2}_{n_0} +O(\beta^{2\kappa_1}) ,  \nonumber\\
\fl p_{\da\ua}^{n_0,\ua}=&\frac{J^2}{4h_{1\op}^{2} } \bigl( \gamma^2 s'^{2}_{n_0}+u'^{2}_{n_0} + O(\beta^{\kappa_1})  \bigr) ,\quad p_{\da\da}^{n_0,\ua}=\frac{J^2}{4h_{1\op}^{2} }\bigl( \gamma^2 t'^{2}_{n_0}+w'^{2}_{n_0}  + O(\beta^{\kappa_1})  \bigr)  . \label{perturbed_form_of_the_other_elements_ua}
\end{eqnarray}

Then, we sum $\{F_1^{n,\ua},F_2^{n,\ua}\}$ and $\{p_{\ua\ua}^{n,\ua},p_{\ua\da}^{n,\ua},p_{\da\ua}^{n,\ua},p_{\da\da}^{n,\ua}\}$ with the Boltzmann weight $e^{-\beta E_{{\rm  tot}}^{'n,\ua}}$ over the label $n$.
First, we calculate the summation of each of $\{F_1^{n,\ua},F_2^{n,\ua}\}$.
From the same discussion as in Eq.~\eqref{Oalpha1} in the case~(b), we have
\begin{eqnarray}
\fl \sum_{n=1}^{2^{N-1}} e^{-\beta E_\md^{'n} } t_n' u'_n  \langle \tilde{\phi}_n^{\ua\da} | \tilde{\phi}_n^{\da\ua} \rangle = O(\beta^{\alpha'}), \quad 
\sum_{n=1}^{2^{N-1}} e^{-\beta  E_\md^{'n}} s_n' w'_n \langle \tilde{\phi}_n^{\da\da} | \tilde{\phi}_n^{\ua\ua} \rangle = O(\beta^{\alpha'}) , \label{Oalpha21}
\end{eqnarray}
where the exponent $\alpha'$ is defined as follows:
\begin{eqnarray}
&\alpha' = \min(\alpha'_1,\alpha'_2) ,\nonumber\\
&\langle \sigma_2^x  \sigma_{N}^x \rangle  = O(\beta ^{\alpha'_1}) ,\quad \langle \sigma_2^y  \sigma_{N}^y \rangle  = O(\beta ^{\alpha'_2}),     
\end{eqnarray}
where $\alpha'_1\ge N-2$ and $\alpha'_2\ge N-2$.
Since we are considering the case $N\ge4$, we have $\alpha'_1\ge2$ and $\alpha'_2\ge2$.
On the other hand, because $\{\delta E_{{\rm  tot}}^{'n,\eta}\}$ in Eq.~\eqref{Eigenvalues_of_total_spin_system2} are of order of $\beta^0$, we have
\begin{eqnarray}
e^{-\beta E_{{\rm  tot}}^{'n,\eta}} = e^{-\beta (E_1^{\eta} + E_{\md}^{'n})} \Bigl(1-\beta  \delta E_{{\rm  tot}}^{'n,\eta} + O(\beta^2) \Bigr) , \label{restriction_weight2}
\end{eqnarray}
for $n=1,2,\cdots,2^{N-1}$ and $\eta=\ua,\da$.
Then, we obtain from Eqs.~\eqref{additive_P_case_c}, \eqref{perturbed_form_of_F2_n0_ua}, and  \eqref{Oalpha21}--\eqref{restriction_weight2},
\begin{eqnarray}
\fl&\sum_{n=1}^{2^{N-1}}  e^{-\beta E_{{\rm  tot}}^{'n,\ua}} F_1^{n_0,\ua}=\sum_{n=1}^{2^{N-1}}  e^{-\beta (E_1^{\ua} + E_{\md}^{'n})}  \Bigl(1-\beta  \delta E_{{\rm  tot}}^{'n,\ua} + O(\beta^2) \Bigr) F_1^{n_0,\ua}  \nonumber\\
\fl=& e^{-\beta E_1^{\ua}} \frac{J}{-2h_{1\op}}  \Bigl( O(\beta^{\alpha'}) + O(\beta^{\kappa_1})+ O(\beta) \Bigr)   =e^{-\beta E_{1}^{\ua}} \times O(\beta^{\kappa_1+1})  
\end{eqnarray}
and 
\begin{eqnarray}
\sum_{n=1}^{2^{N-1}} e^{-\beta E_{{\rm  tot}}^{'n,\ua}} F_2^{n_0,\ua} =e^{-\beta E_{1}^{\ua}} \times O(\beta^{\kappa_1+1})  .
\end{eqnarray}
We similarly calculate the contributions of the other states $\{\ket{\psi_{{\rm  tot}}^{n,\da}}\}$; then, we finally arrive at \eqref{Paraorders21},
\begin{eqnarray}
F_1= \sum_{\eta =\ua,\da}\frac{e^{-\beta E_{1}^{\eta}}}{Z_\tot} \times O(\beta^{\kappa_1+1}) = O(\beta^{\kappa_1+1})   , \nonumber \\
F_2= \sum_{\eta=\ua,\da} \frac{e^{-\beta E_{1}^{\eta}}}{Z_\tot}\times O(\beta^{\kappa_1+1}) =O(\beta^{\kappa_1+1})    ,
\end{eqnarray}
where we utilized the inequality
$
e^{-\beta E_{1}^{\eta}}/Z_\tot <1
$
for $\eta=\ua,\da$.

Second, we calculate the summation of each of $\{p_{\ua\ua}^{n,\ua},p_{\ua\da}^{n,\ua},p_{\da\ua}^{n,\ua},p_{\da\da}^{n,\ua}\}$. 
From Eqs.~\eqref{order_of_max_mixedness2}, \eqref{additive_P_case_c} and \eqref{perturbed_form_of_the_other_elements_ua}, we obtain 
\begin{eqnarray}
\fl p_{\ua\ua}\ge \sum_{n=1}^{2^{N-1}}  \frac{e^{-\beta E_{{\rm  tot}}^{'n,\ua}}}{Z_{{\rm  tot}}}  p_{\ua\ua}^{n,\ua}  = \sum_{n=1}^{2^{N-1}}  \frac{e^{-\beta E_{{\rm  tot}}^{'n,\ua}}}{Z_{{\rm  tot}}}  \bigl(s'^{2}_{n}+u'^{2}_{n}+O(\beta^{2\kappa_1}) \bigr) \ge   W'_{\ua\ua}    
\end{eqnarray}
because $s'^{2}_{n}+u'^{2}_{n}$ cannot vanish for all $n$, where we define the positive number independent of $\beta$ as $W'_{\ua\ua}$ so as to satisfy the above inequalities.
Similarly, for $p_{\ua\da}$, $p_{\da\ua}$ and $p_{\da\da}$, we have
\begin{eqnarray}
\fl&  p_{\ua\da}\ge \sum_{n=1}^{2^{N-1}}  \frac{e^{-\beta E_{{\rm  tot}}^{'n,\ua}} }{Z_{{\rm  tot}}} p_{\ua\da}^{n,\ua} \ge W'_{\ua\da}  ,\quad p_{\da\ua}\ge \sum_{n=1}^{2^{N-1}}  \frac{e^{-\beta E_{{\rm  tot}}^{'n,\ua}} }{Z_{{\rm  tot}}} p_{\da\ua}^{n,\ua} \ge\beta^{2\kappa_1}W'_{\da\ua}         , \nonumber\\
\fl&  p_{\da\da}\ge \sum_{n=1}^{2^{N-1}}  \frac{e^{-\beta E_{{\rm  tot}}^{'n,\ua}}}{Z_{{\rm  tot}}}  p_{\da\da}^{n,\ua} \ge\beta^{2\kappa_1}W'_{\da\da}    .     
\end{eqnarray}
where $W'_{\ua\da}$, $W'_{\da\ua}$ and $W'_{\da\da}$ are positive numbers which do not depend on $\beta$.
We thereby obtain \eqref{Paraorders22} in the limit $\beta\to\infty$.
We thus obtain \eqref{Paraorders21} and \eqref{Paraorders22} and hence the inequality~\eqref{Prove_this_inequality} is satisfied below a certain value of $\beta$ in the case~(c).

Thus, we prove the inequality~\eqref{Prove_this_inequality} in the cases~(a), (b) and (c).
This completes the proof of Theorem~1.

In  the case where the focused two spins are separated by only one spin, Theorem~1 does not apply; the elements $|F_1|$, $\sqrt{p_{\ua\ua}p_{\da\da}}$, $|F_2|$ and $\sqrt{p_{\ua\da}p_{\da\ua}}$ are shown to be of the same order in the same way as in the proof of Theorem~1. 
Therefore, it generally depends on the interaction Hamiltonian and the positions of the focused spins whether the  temperature~$\bar{T}_c$ is finite or not.
As for the three-spin $XYZ$ chains~\eqref{Fundamental_Hamiltonian_Setup} with $\{J_1^x,J_1^y,J_1^z\}=\{J_{2}^x,J_{2}^y,J_{2}^z\}=\{J^x,J^y,J^z\}$ ($J^x\ge J^y\ge J^z$),
we prove in \ref{appendixB_paper3} that the entanglement between the spins~1 and 3 can exist at any temperatures by letting $h_1^z=h_3^z \rightarrow \infty$ except the case of the Ising chain with $h_{2}^z=0$.

%%%%%%%%%%%%%%%%%%%%%%%%%%%%%%%%%%%%%%%%%%%%%%%%%%%%%%%%%%%%%%%%%%%%%%%%%%%%%%%%%%%%%%%%%%%%%%%%%%%%%%%%%%%%%%%%
%%%%%%%%%%%%%%%%%%%%%%%%%%%%%%%%%%%%%%%%%%%%%%%%%%%%%%%%%%%%%%%%%%%%%%%%%%%%%%%%%%%%%%%%%%%%%%%%%%%%%%%%%%%%%%%%
%%%%%%%%%%%%%%%%%%%%%%%%%%%%%%%%%%%%%%%%%%%%%%%%%%%%%%%%%%%%%%%%%%%%%%%%%%%%%%%%%%%%%%%%%%%%%%%%%%%%%%%%%%%%%%%%
%%%%%%%%%%%%%%%%%%%%%%%%%%%%%%%%%%%%%%%%%%%%%%%%%%%%%%%%%%%%%%%%%%%%%%%%%%%%%%%%%%%%%%%%%%%%%%%%%%%%%%%%%%%%%%%%
%%%%%%%%%%%%%%%%%%%%%%%%%%%%%%%%%%%%%%%%%%%%%%%%%%%%%%%%%%%%%%%%%%%%%%%%%%%%%%%%%%%%%%%%%%%%%%%%%%%%%%%%%%%%%%%%
%%%%%%%%%%%%%%%%%%%%%%%%%%%%%%%%%%%%%%%%%%%%%%%%%%%%%%%%%%%%%%%%%%%%%%%%%%%%%%%%%%%%%%%%%%%%%%%%%%%%%%%%%%%%%%%%
%%%%%%%%%%%%%%%%%%%%%%%%%%%%%%%%%%%%%%%%%%%%%%%%%%%%%%%%%%%%%%%%%%%%%%%%%%%%%%%%%%%%%%%%%%%%%%%%%%%%%%%%%%%%%%%%
%%%%%%%%%%%%%%%%%%%%%%%%%%%%%%%%%%%%%%%%%%%%%%%%%%%%%%%%%%%%%%%%%%%%%%%%%%%%%%%%%%%%%%%%%%%%%%%%%%%%%%%%%%%%%%%%

%

%

\section{Numerical demonstration in four-spin chains}
In the present section, we consider the maximization problem in four-spin chains. 
As has been proved in Theorem~1, the four-spin chain is the shortest one in which the end-to-end entanglement cannot be generated in the high-temperature limit.
We mainly discuss the temperature~$\bar{T}_c$ and its dependence on the interaction of the spins.

 \subsection{Numerical results}
In the present section, we consider the \textit{XY} spin chains given by the Hamiltonian
\begin{eqnarray}
\fl H_{{\rm  tot}}=\sum_{i=1}^{3} ( J^x \sigma_i^x   \sigma_{i+1}^x + J^y \sigma_i^y   \sigma_{i+1}^y)  + h_1^z\sigma_1^z +  h_4^z \sigma_4^z +  h_{\md}^z (\sigma_2^z +  \sigma_3^z).  \label{4spinHamiltonian} 
\end{eqnarray} 
We solve the entanglement maximization problem about the focused spins~1 and 4 by varying $h_1^z$ and $h_4^z$ but fixing the temperature $T$ and the external field $h_\md^z$ on the mediator spins~2 and 3.
In order to solve this maximization problem numerically, we used the random search method and the Newton method together.
According to Theorem~1, there always exists a temperature~$\bar{T}_c$ above which the maximized entanglement is exactly equal to zero because the focused spins~1 and 4 are separated by two spins. 
In this case, the temperature $\bar{T}_c$ depends on the value of $h_\md^z$.

 \begin{figure}
\centering
\subfigure[]{
\includegraphics[clip, scale=0.6]{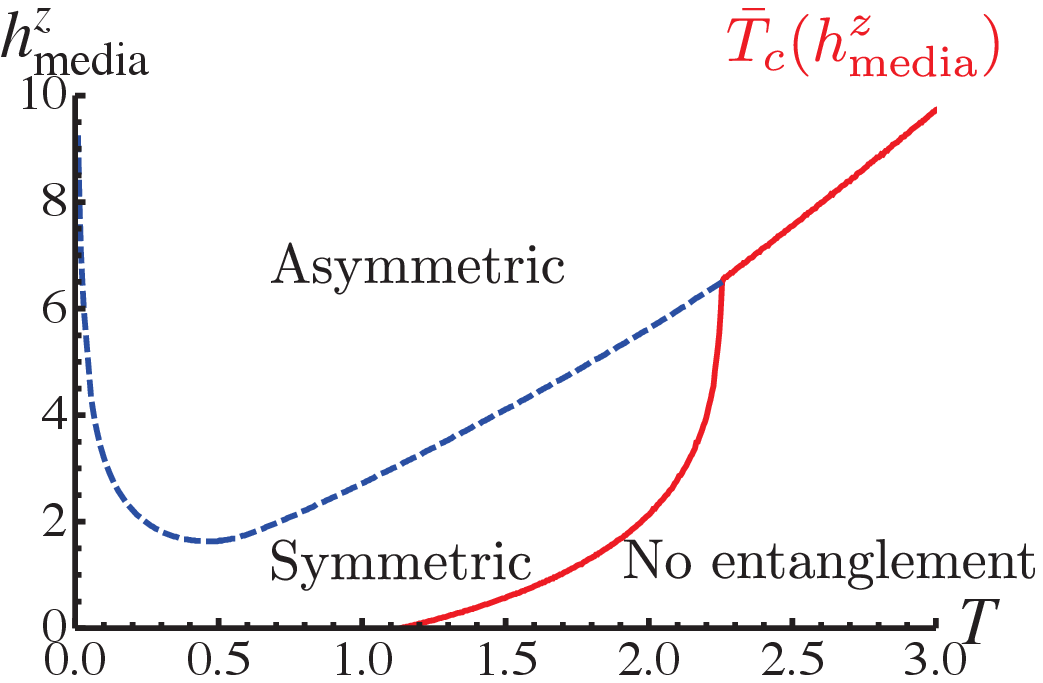}
}
\subfigure[]{
\includegraphics[clip, scale=0.6]{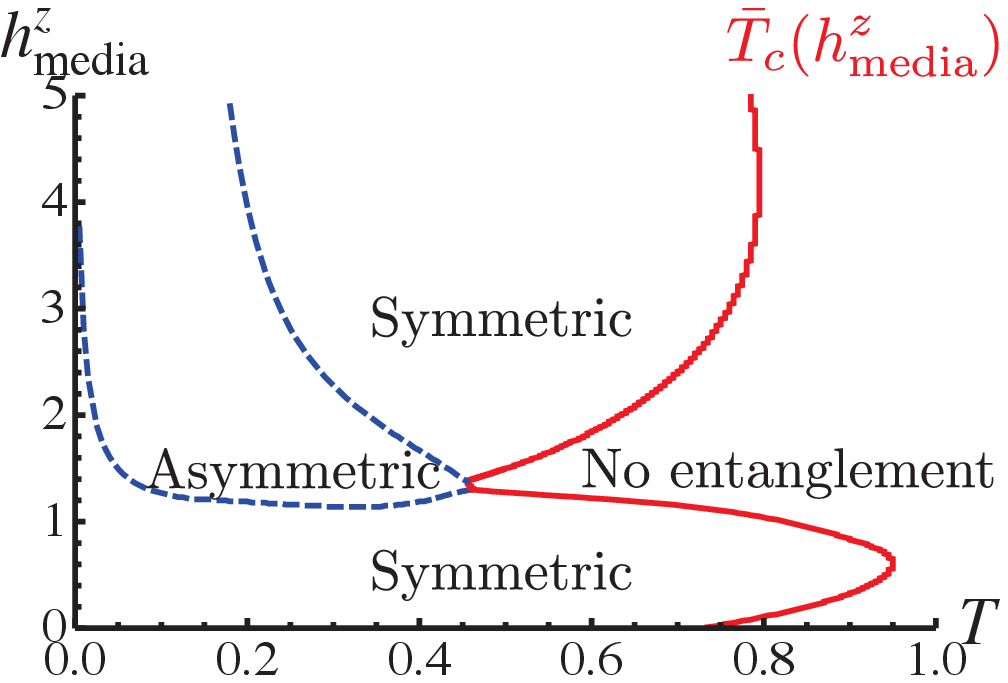}
}
\caption{The phase diagrams for the four-spin chain, (a) in the case of $J^x=J^y=1$ and (b) in the case of $J^x=1,J^y=0.5$.
In the `Asymmetric' phase, the maximizing fields $h_{1\op}$ and $h_{4\op}$ are asymmetric to each other as $|h_{1\op}|\neq |h_{4\op}|$, while they satisfy $|h_{1\op}|= |h_{4\op}|$ in the `Symmetric' phase.
The asymmetry appears on the broken lines.
The maximized entanglement vanishes beyond the solid line, which gives the temperature $\bar{T}_c(h_\md^z)$.}
\label{fig:4spin_XX_Phase}
\end{figure}

We show the phase diagram of the \textit{XX} spin chain and the \textit{XY} spin chain in Fig.~\ref{fig:4spin_XX_Phase}.
In a certain parameter region, there is an `Asymmetric' phase where the maximizing local field $|h_{1\op}|$ is not equal to $|h_{4\op}|$.
The qualitatively different behavior of the temperature~$\bar{T}_c(h_\md^z)$ between the \textit{XX} and the \textit{XY} chains is due to the conservation of the angular momentum in the \textit{z} direction, as we will argue in Section~\ref{argument_asymmetry_section3}.

\subsection{Difference between the \textit{XX} and the \textit{XY} model} \label{argument_asymmetry_section3}
We discuss the behavior of the temperature~$\bar{T}_c$ in the \textit{XX} and \textit{XY} chains.
In the \textit{XX} chain, the temperature~$\bar{T}_c$ increases as the external field $h_\md^z$ is increased, while in the \textit{XY} chain it does not.
This difference is attributed to the conservation of the angular momentum in the \textit{z} direction.
In the \textit{XX} chain, we can suppress the mixture of the states with more than two magnons.
We define a magnon as a spin flip; a down spin in the background of up spins or an up spin in the background of down spins.
For example, the magnon number is two for both of the states $\ket{{\ua\ua\ua\da\da}}$ and $\ket{{\ua\ua\da\da\da}}$.
We choose the fields as
\begin{eqnarray}
&h_1^z= -h_0 ,\nonumber\\
&h_i^z = h_0,\  {\rm  for} \  i=2,3, \cdots N,
\end{eqnarray}
and $h_0\beta \gg1$.
Then, the density matrix $e^{-\beta H_{{\rm  tot}}}$ is almost equivalent to the ground state of $H_{{\rm  tot}}$ and the mixture of the other states is suppressed exponentially by increasing $h_0$.
The ground state is given by the following form;
\begin{eqnarray}
\fl a_0 \ket{{\da_1\ua_2\cdots \ua_N}} + a_1\ket{{\ua_1\da_2\cdots\ua_N}}   + a_2 \ket{{\ua_1\ua_2\da_3 \cdots \ua_N}}+ \cdots + a_{N-1} \ket{{\ua_1\ua_2 \cdots \da_N}} ,\label{Ground_state_4spin_maximization_XX_example}
\end{eqnarray} 
where we calculate $a_k$ by the $k$th order perturbation to have
\begin{eqnarray}
a_k= O\biggl(\frac{J^k}{h_0^{k}} \biggr)  
\end{eqnarray} 
for $k=1,2,\cdots N-1$ with the factor $h_0^k$ coming from the energy denominator.
In the ground state \eqref{Ground_state_4spin_maximization_XX_example}, the element $p_{\da\da}$ is equal to zero because there is no state with more than one down spins in \eqref{Ground_state_4spin_maximization_XX_example}.  
The entanglement between the spins~1 and $N$ exists because $|F_1|\propto|a_{N-1}|\propto|J/h_0|^{N-1}>0$.
The mixture of the excited states generally destroys the entanglement, but is suppressed exponentially because of the Boltzmann weights. 
The entanglement between the spins~1 and $N$ thereby survives.

On the other hand, in the \textit{XY} chains, we cannot control the number of the magnons in the ground state by increasing $h_\md^z$.
Therefore,  $p_{\ua\ua}p_{\da\da}$ is not zero in the ground state, which invalidates the argument for the \textit{XX} model.
This may account for the fact that the temperature~$\bar{T}_c$ does not increase as the external field $h_\md^z$ is increased.

\section{Summary and conclusion}\label{conclusion}
In this paper, we have analytically investigated the increase of the critical temperature by the local fields in systems with more than two spins.
For this problem, we had two possibilities: in the high-temperature limit, the entanglement completely vanishes or infinitesimally remains. 
In order to study it, we have introduced the maximum value of the entanglement between the focused two spins under the condition that we can arbitrarily tune the local fields only on these two spins.
We have shown the general theorem on its spin-number dependence.
We proved that the maximized entanglement is equal to zero above a certain temperature $\bar{T}_c$ if the two spins are separated by two or more spins.
It means that the critical temperature $T_c$ cannot be enhanced beyond $\bar{T}_c$ by the local fields.
This result tells us that as the temperature goes higher the thermal entanglement between distantly separated two spins cannot survive with any local controls.
On the other hand, the entanglement follows a  power law decay of the temperature if the two spins are separated by one spin.

In the four-spin chains, we have numerically demonstrated that the entanglement vanishes above a certain temperature $\bar{T}_c$ even after the maximization.
Because of the difference in the symmetric property between the \textit{XX} and \textit{XY} spin chains, the dependence of the temperature~$\bar{T}_c$ on the fields $h_{\md}^z$ are qualitatively different between the two systems.

In conclusion, our study has given one of the general limits for the entanglement generation.
We have investigated the effect of the purity increase by the local fields on the entanglement generation.
However, there are other problems to be solved. 
For example, we have not considered the case where we can arbitrarily modulate the external fields not only on the focused spins but also on the mediator spins.
In this case, we may also have a similar theorem to Theorem~1 though the calculation of the entanglement will be much more complicated.
As another problem, we have not discussed the influence of the external fields on the global entanglement, which is of more interest to many researchers.
In future, we plan to investigate such a unsolved problem.

\section*{ACKNOWLEDGMENT}
The present author is grateful to Professor Naomichi Hatano for helpful discussions and comments.
%The present study is supported by CREST from Japan Science and Technology Agency as well as Grant-in-Aid for scientific Research No.~22340110.

\appendix

\section{The case $\bf{\tilde{\kappa}\le0}$ in \eqref{Paraorders1_for_Appendix}} \label{appendixE_paper3}
Here, we discuss the case of $\tilde{\kappa}\le0$ in the case~(b).
In this case, we cannot consider the unperturbed states $\ket{{\ua_1\da_N}}\otimes\ket{\psi_{{\rm  media}}^n}$ and $\ket{{\da_1\ua_N}}\otimes\ket{\psi_{{\rm  media}}^n}$ independently because their eigenvalues are almost degenerate.
As a result, the parameter $\kappa'=\min(\kappa_N,\tilde{\kappa},1)$ does not satisfy the inequality~\eqref{Definition_of_the_two_kappa}.
We can still apply the same calculation to the other parameters $\{p_{\ua\da}p_{\da\ua},p_{\ua\ua}p_{\da\da}\}$ as in the case $\tilde{\kappa}>0$; they are of order of $\beta^{2\kappa_1+2\kappa_N }$ or lower. 

We here prove that $F_1^2$ and $F_2^2$ are of order higher than $\beta^{2\kappa_1+2\kappa_N }$.
In order to prove this, we separate $H_1+H_N$ as follows by letting $ \tilde{h}_{1\op}\equiv ( h_{1\op}+h_{N\op}+ h_0)/2$ and $ \tilde{h}_{N\op}\equiv ( h_{1\op}+h_{N\op}- h_0)/2$;
\begin{eqnarray}
&\fl H_1+H_N = \tilde{H}_{{\rm  LO}} + \delta H_{{\rm  LO}}, \nonumber\\
&\fl \tilde{H}_{{\rm  LO}}=\tilde{h}_{1\op} \sigma_1^z +\tilde{h}_{N\op}  \sigma_N^z =-(h_{1\op}+h_{N\op}) \bigl ( \ket{{\ua_1\ua_N}} \bra{{\ua_1\ua_N}} -\ket{{\da_1\da_N}} \bra{{\da_1\da_N}}\bigr ) \nonumber\\
&\fl \quad \quad \quad \quad \quad \quad\quad \quad \quad \quad \quad \quad - h_0 \bigl( \ket{{\ua_1\da_N}}\bra{{\ua_1\da_N}} - \ket{{\da_1\ua_N}} \bra{{\da_1\ua_N}} \bigl)   ,  \nonumber\\
 &\fl \delta H_{{\rm  LO}}= (h_{1\op}-\tilde{h}_{1\op}) \sigma_1^z+(h_{N\op}-\tilde{h}_{N\op})\sigma_N^z  \nonumber\\
&\fl \quad \quad \ \ = - (h_{1\op}-h_{N\op}-h_0) \bigl( \ket{{\ua_1\da_N}}\bra{{\ua_1\da_N}} - \ket{{\da_1\ua_N}}\bra{{\da_1\ua_N}} \bigl)  ,
\end{eqnarray}
where we define as $h_0\equiv W_0 \beta^{-\tilde{\kappa}_0}$ with $0<\tilde{\kappa}_0<1$ and $W_0>0$, and regard $\beta \delta H_{{\rm  LO}}$ as perturbation.
The Hamiltonian $\tilde{H}_{{\rm  LO}}$ satisfies the inequality \eqref{Paraorders1_for_Appendix}  because $\tilde{h}_{1\op}-\tilde{h}_{N\op}=W_0 \beta^{-\tilde{\kappa}_0}$ is of order of $\beta^{-\tilde{\kappa}_0}$ with $0<\tilde{\kappa}_0<1$.
Then, the unperturbed density matrix $\tilde{\rho}_{{\rm  tot}}^{(0)}$ is given by
\begin{eqnarray}
\tilde{\rho}_{{\rm  tot}}^{(0)}& =  e^{-\beta \tilde{H}_{{\rm  tot}}^{(0)} }, \quad 
\tilde{H}_{{\rm  tot}}^{(0)} &=H_{{\rm  couple}}+ H_{{\rm  media}} +\tilde{H}_{{\rm  LO}},
\end{eqnarray} 
where $H_{{\rm  media}}$ and $H_{{\rm  couple}}$ are given in \eqref{Definition_H_media_couple}.
The magnitudes of the unperturbed elements $\{F_1^{(0)}, F_2^{(0)}\}$ of $\tilde{\rho}_{{\rm  tot}}^{(0)}$ are given by \eqref{Paraorders1}, namely,
\begin{eqnarray}
\beta^{\kappa_1+\kappa_N + \kappa'_0}  \ {\rm  and} \ \beta^{\kappa_1+\kappa_N + \kappa_0}  \ {\rm  or}\ {\rm  higher} \label{F_01_order_appendix}
\end{eqnarray}
where
$
\kappa_0=\min(\kappa_N,1) \ {\rm  and}\  \kappa'_0=\min(\kappa_N,\tilde{\kappa}_0,1) .
$

The density matrix in the first-order perturbation is given by
\begin{eqnarray}
\fl\rho_{{\rm  tot}}=\frac{1}{Z^{(0)}+\delta Z}\biggl( e^{-\beta \tilde{H}_{{\rm  tot}}^{(0)} } +\beta \int_0^1  e^{-\beta x \tilde{H}_{{\rm  tot}}^{(0)} } \delta H_{{\rm  LO}} e^{-\beta (1-x) \tilde{H}_{{\rm  tot}}^{(0)}}  dx \biggr),
\end{eqnarray}
where $Z^{(0)}$ is the partition function of $e^{-\beta \tilde{H}_{{\rm  tot}}^{(0)}}$, while $Z^{(0)}+\delta Z$ is the partition function of $e^{-\beta H_{{\rm  tot}}}$.
Now, we calculate the elements 
$
F_1= \tr_{1N}  \bra{{\ua_1\da_N}} \rho_{{\rm  tot}} \ket{{\da_1\ua_N}},
$
where $\tr_{1N}$ denotes the trace operation on the spins except the focused spins $1$ and $N$.
The first-order perturbations of the elements $F_1$ is
\begin{eqnarray}
\fl\frac{-\beta (h_{1\op}-h_{N\op}-h_0)}{Z^{(0)}+\delta Z}  \tr_{1N}& \int_0^1 dx \biggl(  \bra{{\ua_1\da_N}}  e^{-\beta x \tilde{H}_{{\rm  tot}}^{(0)}} \ket{{\ua_1\da_N}}\bra{{\ua_1\da_N}} e^{-\beta (1-x) \tilde{H}_{{\rm  tot}}^{(0)} }  \ket{{\da_1\ua_N}}\nonumber\\
\fl&- \bra{{\ua_1\da_N}}  e^{-\beta x \tilde{H}_{{\rm  tot}}^{(0)}} \ket{{\da_1\ua_N}} \bra{{\da_1\ua_N}} e^{-\beta (1-x) \tilde{H}_{{\rm  tot}}^{(0)} } \ket{{\da_1\ua_N}} \biggr) .\label{F_1_variation_appendix}
\end{eqnarray}

In order to estimate the order of the first-order perturbations of $F_1$, we introduce $\{\bket{\psi_{{\rm  tot},\xi}^{(0),n,\xi}}\}$ and $\{E_{{\rm  tot}}^{(0),n,\xi}\}$ as defined in Eqs.~\eqref{Eigenstates_of_total_spin_system} and \eqref{Eigenvalues_of_total_spin_system}, but for $\tilde{H}_{{\rm  tot}}^{(0)}$.
First, we have 
\begin{eqnarray}
\fl \bra{{\ua_1\da_N}} e^{-\beta x \tilde{H}_{{\rm  tot}}^{(0)} }  \ket{{\da_1\ua_N}}=\sum_{n=1}^{2^{N-2}}  \sum_{\xi=\ua\ua,\ua\da,\da\ua,\da\da} e^{-\beta x E_{{\rm  tot}}^{(0),n,\xi}} \bket{\psi_{{\rm  media},\ua\da}^{(0),n,\xi}} \bbra{ \psi_{{\rm  media},\da\ua}^{(0),n,\xi} } .
\end{eqnarray}
From the calculation in section~1 of the supplementary materials, we obtain
\begin{eqnarray}
\Bigl |\Bigl | \bket{\psi_{{\rm  media},\ua\da}^{(0),n,\xi}} \bbra{ \psi_{{\rm  media},\da\ua}^{(0),n,\xi} } \Bigr| \Bigr| = O(\beta^{\kappa_1+\kappa_N}).
\end{eqnarray}
Therefore, we have
\begin{eqnarray}
 \bigl |\bigl | \bra{{\ua_1\da_N}}\frac{e^{-\beta x \tilde{H}_{{\rm  tot}}^{(0)} } }{Z^{(0)}(x)}  \ket{{\da_1\ua_N}} \bigr| \bigr| = O(\beta^{\kappa_1+\kappa_N}),
\end{eqnarray}
where 
$
Z^{(0)}(x) \equiv \tr \bigl(e^{-\beta x \tilde{H}_{{\rm  tot}}^{(0)}} \bigr).
$
As a result, we obtain the first-order perturbations of $F_1$ as
\begin{eqnarray}
 O(\beta^{\kappa_1+\kappa_N +1-\tilde{\kappa}_0}),  \label{F_1_variation_first_order_appendix}
\end{eqnarray}
where we utilized $\beta (h_{1\op}-h_{N\op}-h_0) = O(\beta^{1-\tilde{\kappa}_0})$ and 
\begin{eqnarray}
\frac{Z^{(0)}(x)Z^{(0)}(1-x)}{Z^{(0)}+\delta Z} = O(\beta^{0}).
\end{eqnarray}
We can similarly calculate higher-order perturbations of $F_1$ to see that $F_1$ is of order higher than \eqref{F_1_variation_first_order_appendix}.
By the same calculation, we can also prove that $F_2$ is of order higher than $\beta^{2\kappa_1+2\kappa_N }$.
Thus, in the case $\tilde{\kappa}\le0$, $F_1^2$ and $F_2^2$ are of order higher than $\{p_{\ua\da}p_{\da\ua},p_{\ua\ua}p_{\da\da}\}$, which are of order of $\beta^{2\kappa_1+2\kappa_N }$ or lower.

\section{Proof for the existence of the entanglement in three-spin systems} \label{appendixB_paper3}
In this section, we prove the statement in Section~4 that the maximized entanglement always exists in the systems~\eqref{Fundamental_Hamiltonian_Setup} with $\{J_1^x,J_1^y,J_1^z\}=\{J_{2}^x,J_{2}^y,J_{2}^z\}=\{J^x,J^y,J^z\}$ $(J^x\ge J^y\ge J^z)$ in the high-temperature limit $\beta\rightarrow0$, except the case of the Ising chain with $h_{2}^z=0$.
We prove it by showing that the entanglement exists by letting $h_1^z$ and $h_3^z$
\begin{eqnarray}
h_1^z=h_3^z=h_0(\beta) ,
\end{eqnarray}
where we define 
$
h_0(\beta)  \equiv W_0 \beta^{-\kappa_0}
$
with $\kappa_0>1$  and  $W_0>0$.
We regard the interaction Hamiltonian as perturbation and calculate the leading order of the elements $\{F_1,F_2,p_{\ua\ua},p_{\ua\da},p_{\da\ua},p_{\da\da}\}$ of the density matrix.
Then we show the inequality~\eqref{sufneccond1_paper3}, which is a necessary and sufficient condition for the existence of the entanglement.

First, we separate the Hamiltonian~\eqref{Fundamental_Hamiltonian_Setup} as follows:
\begin{eqnarray}
H_{{\rm tot}} = \tilde{H}_{{\rm LO}}  +\tilde{H}_{{\rm couple}},
\end{eqnarray} 
where 
\begin{eqnarray}
\fl&\tilde{H}_{{\rm LO}}=  h_1^z \sigma_1^z +  h_3^z \sigma_3^z +  h_{2}^z \sigma_2^z + J^z \sigma_1^z   \sigma_2^z+ J^z \sigma_2^z   \sigma_{3}^z ,  \nonumber \\
\fl&\tilde{H}_{{\rm couple}}=(J^x \sigma_1^x   \sigma_{2}^x + J^y \sigma_1^y   \sigma_{2}^y) +(J^x \sigma_2^x   \sigma_{3}^x + J^y \sigma_2^y   \sigma_{3}^y)  \nonumber \\
\fl               &\quad \ =J\bigl[\sigma_1^+   \sigma_{2}^- + \sigma_1^-   \sigma_{2}^+  +\sigma_{2}^+   \sigma_{3}^- + \sigma_{2}^-   \sigma_{3}^+  + \gamma (\sigma_1^+   \sigma_{2}^+ + \sigma_1^-   \sigma_{2}^-+\sigma_{2}^+   \sigma_{3}^+ + \sigma_{2}^-   \sigma_{3}^-) \bigr] ,
\end{eqnarray} 
$h_{2}^z$ has an arbitrary value, $J^x=J(1+\gamma)/2$, $J^y=J(1-\gamma)/2$, and we consider $\tilde{H}_{{\rm int}}$ as perturbation. 
Because $\beta h_0(\beta)\rightarrow\infty$ as $\beta\rightarrow0$, we only have to consider the ground state and the first excited state as the unperturbed states, which are given by
$
\ket{{\ua_1\ua_2\ua_3}}\   {\rm and} \   \ket{{\ua_1\da_2 \ua_3}} 
$
with the corresponding eigenvalues
\begin{eqnarray}
\epsilon_1 = -2h_0 -h_{2}^z + 2J^z\   {\rm and} \ \epsilon_2 = -2h_0 +h_{2}^z -2J^z  ,
\end{eqnarray}
where we assume $h_{2}^z>0$, but the following discussion is also applicable to the case of $h_{2}^z<0$.  
The second excited states have the eigenvalues $\epsilon_1+2h_0 +O(\beta^0)$ and hence its thermal mixing can be ignored in the limit $\beta\rightarrow 0$; similarly, the mixing of the higher excited states can be also ignored.
By mixing these two states with the Boltzmann weights $e^{-\beta \epsilon_1}$ and $e^{-\beta \epsilon_2}$, we obtain the matrix elements $\{F_1,F_2,p_{\ua\ua},p_{\ua\da},p_{\da\ua},p_{\da\da}\}$.

Let us first consider the case $\gamma\neq1$.
In this case, the leading terms of $F_1$ and $\sqrt{p_{\ua\ua} p_{\da\da}}$ are given as follows, respectively:
\begin{eqnarray}
&F_1 = \frac{J^2(1+\gamma^2)}{8 h_0^2}  + O(\beta^{1+2\kappa_0}),  \quad \sqrt{p_{\ua\ua} p_{\da\da}} =\frac{J^2 \gamma}{4 h_0^2}+ O(\beta^{1+2\kappa_0})  ,
 \end{eqnarray}
which yields
\begin{eqnarray}
&F_1-\sqrt{p_{\ua\ua} p_{\da\da}} =\frac{J^2(1-\gamma)^2}{8 h_0^2} +  O(\beta^{1+2\kappa_0})  > 0  \label{con3spinproof}
 \end{eqnarray}
 in the limit $\beta\rightarrow0$.
This gives a non-zero value of the concurrence~\eqref{conexpression} of order of $\beta^{2\kappa_0}$ in the case $\gamma\neq0$.

Next, we consider the case $\gamma=1$.
Because we assumed $J^x\ge J^y\ge J^z$, the equality $J^y=J^z=0$ is satisfied in the case $\gamma=1$.
In this case, we have to take higher-order approximation because the first term of Eq.~\eqref{con3spinproof} vanishes.
The expansions of $F_2$ and $\sqrt{p_{\ua\da} p_{\da\ua}}$ are given, respectively, by
\begin{eqnarray}
\fl&F_2 =\frac{J^2}{4 h_0^2}  +\frac{(h_2^{z})^2 J^2}{8} \frac{\beta}{h_0^2}   -\frac{(h_2^{z})^2 J^2}{4} \frac{\beta}{h_0^3}   + O(\beta^{2+3\kappa_0})  ,\nonumber \\
\fl&\sqrt{p_{\da\ua} p_{\ua\da}} = \frac{J^2}{4 h_0^2}  +\frac{(h_2^{z})^2 J^2}{8} \frac{\beta}{h_0^2}   -\frac{(h_2^{z})^2 J^2}{2}\frac{\beta}{h_0^3}   +O(\beta^{2+3\kappa_0}) ,
\end{eqnarray}
which is followed by
\begin{eqnarray}
F_2-\sqrt{p_{\ua\da} p_{\da\ua}} =\frac{\beta}{4h_0^3} J^2 (h_2^{z})^2+  O(\beta^{2+3\kappa_0}) .
\end{eqnarray}
If the field $h_2^z$ is equal to zero, the entanglement vanishes for any values of the local fields $h_1^z$ and $h_3^z$; it is because in this case the interaction is classified into `classical' interaction~\cite{Kuwahara3}.
For $h_2^z\neq0$, the concurrence~\eqref{conexpression} is of order of $\beta^{1+3\kappa_0} $ in the case $\gamma=1$ and increases as the field $h_2^z$ is increased. 
We have thereby proved that the entanglement in a three-spin system~\eqref{Fundamental_Hamiltonian_Setup} with $\{J_1^x,J_1^y,J_1^z\}=\{J_{2}^x,J_{2}^y,J_{2}^z\}=\{J^x,J^y,J^z\}$ $(J^x\ge J^y\ge J^z)$ has always a non-zero value except the case of the Ising chain with $h_{2}^z=0$ if we choose the local fields properly.

%

%
%

%

%%%%%%%%%%%%%%%%%%%%%%%%%%%%%%%%%%%%%%%%%%%%%%%%%%%%%%%%%%%%%%%%%%%%%%%%%%%%%%%%%%%%%%%%%%%%%%%%%%%%%%%%%%%%%%%%
%%%%%%%%%%%%%%%%%%%%%%%%%%%%%%%%%%%%%%%%%%%%%%%%%%%%%%%%%%%%%%%%%%%%%%%%%%%%%%%%%%%%%%%%%%%%%%%%%%%%%%%%%%%%%%%%
%%%%%%%%%%%%%%%%%%%%%%%%%%%%%%%%%%%%%%%%%%%%%%%%%%%%%%%%%%%%%%%%%%%%%%%%%%%%%%%%%%%%%%%%%%%%%%%%%%%%%%%%%%%%%%%%
%%%%%%%%%%%%%%%%%%%%%%%%%%%%%%%%%%%%%%%%%%%%%%%%%%%%%%%%%%%%%%%%%%%%%%%%%%%%%%%%%%%%%%%%%%%%%%%%%%%%%%%%%%%%%%%%
%%%%%%%%%%%%%%%%%%%%%%%%%%%%%%%%%%%%%%%%%%%%%%%%%%%%%%%%%%%%%%%%%%%%%%%%%%%%%%%%%%%%%%%%%%%%%%%%%%%%%%%%%%%%%%%%

%

%

%

%
%

\section*{References}

\renewcommand{\refname}{\vspace{-1cm}}

%\appendix


\begin{thebibliography}{00}
\bibitem{Density_matrix_expansion} 
Because the maximizing local fields $h_{1\op}(\beta)$ and $h_{N\op}(\beta)$ depend on $\beta$, the total Hamiltonian $H_{\tot}$ also depends on $\beta$.
This makes it difficult to calculate the density matrix $e^{-\beta H_{\tot}}$ simply by the use of matrix expansion such as $e^{-\beta H_{\tot}}= I_\tot-\beta H_{\tot}+\beta^2 H_{\tot}^2/2+\cdots$.
Instead, we would have to apply the Dyson expansion, but it would not simplify the proof significantly.
\end{thebibliography}
\end{document}